\documentclass[aps,twocolumn,showpacs,prb,floatfix]{revtex4}
\usepackage{amsmath}
\usepackage{graphicx}
\usepackage{epsfig}
\newlength{\bxwidth}\bxwidth=1.5 truein
\newlength{\upit}\upit=0.1truein
\newcommand{\raiser}[1]{\raisebox{\upit}[0cm][0cm]{#1}}
\newcommand{\ltappr}{{{\lower4pt\hbox{$<$} } \atop \widetilde{ \ \ \
}}}
\newcommand\frmup[1]{\raiser{\epsfig{file=#1,width=\bxwidth}}}

\newcommand{\cg}{{\cal G}}

\newcommand{\Str}{\underline{\hbox{Str}}}

\newcommand{\Tr}{\underline{\hbox{Tr}}}
\newcommand{\dg}{^{\dagger }}
\newcommand{\vk}{\vec k}

\newcommand{\rarrow}{\rightarrow}

\newlength{\figwidth}
\newlength{\shift}
\newcommand{\fg}[3]
{
\begin{figure}[ht]

\vspace*{-0cm}
\[
\includegraphics[width=\figwidth]{#1}
\]
\vskip -0.2cm
\caption{\label{#2}
\small#3
}
\end{figure}}

\newcommand \bea {\begin{eqnarray} }
\newcommand \eea {\end{eqnarray}}
\newcommand{\bk}{{\bf{k}}}

\def\gtappr{{{\lower4pt\hbox{$>$} } \atop \widetilde{ \ \ \ }}}
\begin{document}
%  Working Title:
%  
%  Conservation laws and the emergence of heavy quasiparticles in
%  the Kondo and infinite U Anderson model.

\title{
Fermi liquid identities for the Infinite U multichannel Anderson Model
}
%  

%  Bringing back the electron in a system of spin-charge decoupled
%  gauge fields.
%  
%  
\author{Eran Lebanon and P. Coleman}
\affiliation{
 Center for Materials Theory, Serin Physics Laboratory,
    Rutgers University, Piscataway, New Jersey 08854-8019, USA. }

%\date{}
\begin{abstract}
We show how the electron gas methods of Luttinger, Ward and Nozi\`eres
can be applied to an $SU (N)$, multi-channel generalization of the 
infinite U Anderson impurity model within a
Schwinger boson treatment.  Working to all orders in a $1/N$ expansion,
we show how the Friedel Langreth relationship, the
Yamada-Yosida-Yoshimori
 and the Shiba-Korringa relations can be
derived, under the assumption that the spinon and holon fields
are gapped. One of the remarkable features of this
treatment, is that the Landau amplitudes depend 
on the exchange of low energy virtual spinons and
holons. We end the paper with a discussion on the extension of our
approach to the lattice, where the spinon-holon gap is expected to close
at a quantum critical point.
\end{abstract}

%\eject
%
\maketitle
%
%\vfill\eject 

\section{Introduction}\label{}

The standard model of interacting  electronic systems is based on
the theoretical framework established by Landau\cite{landaufl} in the 1950's, whereby
the properties of interacting fluids of fermions are 
linked to non-interacting fermions by 
adiabatically turning on the interactions. This is the basis of 
the Landau Fermi liquid, and it leads to the concept of quasiparticles.
It also provides the formal 
basis for the diagrammatic description of interacting
Fermi fluids. In the early 1960's, Luttinger and Ward\cite{LW} showed how
many aspects of the Landau Fermi liquid theory - such as the relationship
between fermion density and Fermi surface volume, the relationship
between the linear specific heat capacity and the quasiparticle
density of states, could be deduced from first principle 
resummations of diagrammatic
perturbation theory to infinite order. 

Many new electronic materials discovered over the past two
decades, such as the high temperature superconductors, colossal
magneto-resistance and heavy electron materials, do not fit so
naturally into the Landau scheme.  In these 
materials, the interactions between the fermions become so
large  they project out large tracts of the many body Hilbert space.  
For example, in many
narrow band transition metal or f-electron materials, the charge
fluctuations of the localized orbitals are restricted to certain
valence states, and in local moment systems, the electrons in a
localized moment are restricted to states of fixed definite
occupancy. In these systems, the validity of the adiabatic approach
is increasingly in question, 
and a more appropriate starting point is one where 
the Coulomb interaction has been taken to infinity from the outset,
explicitly removing certain states from the Hilbert space. 
For example, in the 
infinite $U$ Anderson model, states of double occupancy are excluded
whereas in the Kondo model, all charge
fluctuations of the local moment are excluded, as illustrated in
Fig. 1.

\figwidth=\columnwidth
\fg{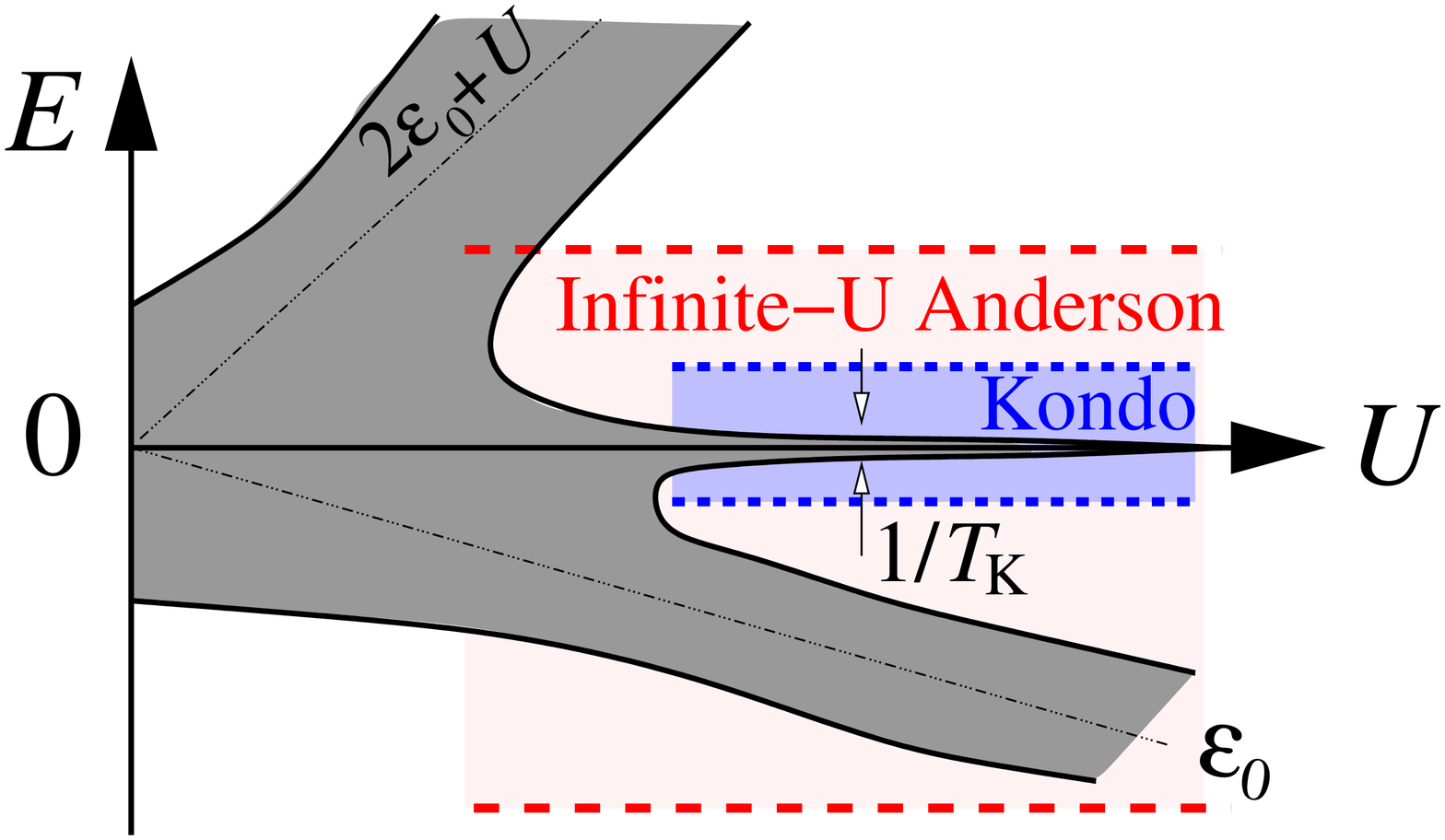}{figmodel}
{Schematic diagram illustrating the evolution of the impurity
f-spectral function with the strength of the repulsive interaction $U$.
Shaded ares denote the regions
where the impurity f-spectral function is large. As $U$ is increased,
high-energy regions of the Hilbert space are projected out of the
low-energy
model, giving rise to the infinite-$U$ Anderson model, and Kondo model
as effective low energy theories.  So long as adiabaticity is
preserved, the spectral weight at the Fermi surface is preserved, no matter how large
$U$ becomes. 
}

When cast as a field theory, infinite U models
become gauge theories in which the Hubbard operators that create and
destroy the highly correlated electrons are written 
as composite products of ``slave
operators''. In such an approach,  
the constraints on charge fluctuations become
conserved local quantities associated with local gauge invariances and 
the basic fields entering into the Hamiltonian are
``fractionalized'' fields, which carry either spin or charge, but not
both.  The gauge theory approach to strongly correlated electron systems 
poses a number of important technical and qualitative
challenges.  In particular, 
{\sl how does 
the Landau Fermi liquid emerge within a Gauge theory of unbroken symmetry, and 
what happens to the spectrum of gauge particle excitations ?}
This question assumes an increased importance in the context of recent speculations
that fractional particles may become free at a quantum critical
point\cite{coleman,pepin05,deconfinedcrita,senthil}.

The last ten years have seen  considerable progress 
in methods that combine the Luttinger Ward approach with a new
generation of  large $N$ expansions 
which use Schwinger bosons,rather
than Abrikosov pseudo-fermions to represent the localized
moments \cite{coxruck93,PG97,indranil05,rech,lebanon}.  
We now briefly review some of these recent developments.
Cox and Ruckenstein were the first to apply a large $N$ approxach to a
multi-channel Kondo single impurity Kondo model, using a fermionic
spin represenation.
Parcollet and Georges[\onlinecite{PG97}] developed the original formalism
for the Schwinger Boson  large $N$ approach to this model, employing a  multi-channel Kondo
model, focusing their interest mainly on the overscreened case
[\onlinecite{indranil05}], built upon this earlier work to develop 
a Luttinger Ward approach, showing how it can be used to derive a
Friedel sum rule for the single impurity model and a 
Luttinger sum rule for the lattice, assuming that the Fermi liquid
state could be treated in a $1/N$ expansion. 
Rech et al. [\onlinecite{rech}] showed how
the fully screened Kondo impurity model could indeed be treated in the large
$N$ limit using the Schwinger boson method, demonstrating the
emergence of Fermi liquid behavior. Lenanon et al. [\onlinecite{lebanon}] extended these
methods to the infinite $U$ Anderson model, showing how a Baym
Kadanoff approximation scheme
can be developed from the  $1/N$ expansion of the Luttinger Ward
functional, using the leading order term to 
develop a conserving
approximation at finite $N$. 

One of the key insights emerging from the large $N$ solutions 
to the fully screened Kondo and Anderson impurity models, is that 
the development of a Fermi liquid 
is accompanied by the formation of a gap $\Delta_{g}$ of order the
Kondo temperature $T_{K}$, in the spectrum of the
fractional particles \cite{rech,lebanon}. 

In this paper we explore the general consequences of this
observation for the infinite $U$ Anderson impurity model.
In particular, we show that the 
the  assumption of
a gap in the fractionalized particle spectrum enables us 
to extend the Luttinger Ward approach  beyond the leading orders
considered so far, to all orders in the $1/N$ expansion.
In our approach, the expansion in powers of $1/N$ plays a
completely analogous role to the expansion in  powers of the
interaction. The Luttinger Ward approach was first applied to the finite $U$ Anderson model
by Yamada, Yosida, Shiba and Yoshimori,\cite{yamada,yoshimori,shiba} who used  conservation laws to
develop a set of conserving identities that apply to all orders in the
strength of the interaction $U$. Here, taking advantage of these new
insights, we show how a parallel set of
results can be obtained for the {\sl infinite} $U$ Anderson model.

Many structural aspects of our work, including the interplay 
between virtual fractional excitations and the composite heavy
quasiparticles may enjoy a wider application to the  Anderson and Kondo
lattices. These are points that we return to in a discussion at the
end of the paper. 

\section{Summary of key aspects of the Paper. }\label{}

The starting point for our work is the infinite $U$ Anderson model,
which we formulate by representing the Hubbard operators 
as a product of slave fermion (``holon'') and a Schwinger boson
(``spinon'' )\cite{jayaprakash89,yoshioka89} $X_{\alpha 0}  = b\dg _{\alpha} \chi$
where $b\dg _{\alpha }$ creates a ``spinon'' with spin component
$\alpha $ and $\chi \dg $ creates a charged ``holon''. 
Since there is no natural perturbative expansion in terms of the
parameters of the original model, we consider a family 
infinite $U$ impurity Anderson models with  $SU (N)$ spin symmetry, 
employing  a $1/N$ expansion around the
large $N$ limit. There are a number of special tricks that need to be
carried out in order to realize such a large $N$ expansion
\begin{itemize}
\item The spin $S$ of the moment, described by the number of Schwinger bosons
$n_{b}=2S$ must be allowed to grow with $N$, so that $2S/N$ remains
finite.

\item  More surprisingly, the number of electron scattering channels $K$
must also grow with $N$\cite{PG97}, in order to screen the large moment $S$.
The special case $2S=K$ describes the fully screened model\cite{rech}.

\end{itemize}
The introduction of $K$ channels requires that we generalize the
Hubbard operators as follows\cite{gan}
\begin{equation}\label{}
X_{\alpha 0}  \rightarrow X_{\alpha \nu }= b\dg _{\alpha} \chi_{\nu},
\end{equation}
where $\nu\in [1,K]$ is the channel index. $X_{\alpha \nu}$ creates a
localized electron with spin component $\alpha $ and channel index $\nu$.

We formulate these features into an infinite $U$ Anderson
impurity model 
\cite{lebanon} as follows, 
\begin{eqnarray}\label{lattice}
{\cal H} &=& \sum_{{\vec k}\nu \alpha} \epsilon_{\vec k}
c^{\dagger}_{{\vec k}\nu \alpha } c_{{\vec k}\nu \alpha}+H_{I} .
\end{eqnarray}
Here $c^{\dagger}_{{\vec k}\nu \alpha}$ creates a conduction
electron with momentum ${\vec k}$, channel index $\nu \in [1,K]$, 
and spin index $\alpha \in [1,N]$.  
The term 
\begin{equation}
H_{I}  =
\frac{{V}}{\sqrt{N}}\sum_{{\vec k}\nu \alpha}
\left[ c^{\dagger}_{{\vec k}\nu \alpha} \chi^{\dagger}_{\nu} 
b_{\alpha}
+{\rm H.c} \right] \nonumber 
+ \epsilon_0 \sum_{\alpha} b^{\dagger}_{\alpha} b_{\alpha}
\label{model}
\end{equation}
describes the interaction of the conduction electrons with a magnetic
ion located at the origin.
The $\sqrt{N}$ denominator in the hybridization ensures 
a well-defined large $N$ limit. The energy of a singly occupied 
impurity $\epsilon_0$ is taken to be negative. The conserved operator 
$Q= \sum_{\alpha } b_{\alpha }^{\dagger}b_{\alpha } + \sum_{\nu} 
\chi^{\dagger}_{\nu } \chi_{\nu}$ generalizes  the no-double occupancy 
constraint of the infinite $U$ 
Anderson model, by restricting the ``valence'' $n_{b}\leq  Q$.  
$Q$ is also the maximum size $S= Q/2$ of the local moment that
can develop at each site. The condition  $Q=K$ is 
required for perfect screening of the local moment. 

The spinon and holon operators carry the conserved charge $Q$,
and in order to discuss these fields, we need to consider the enlarged
Fock space of general $Q\neq K$. 
The partition
function for the general problem is given by
\[
Z_{Q} = {\rm  Tr} \left[{\cal P}_{Q}e^{-\beta H} \right],
\]
where the projection operator ${\cal P}_{Q}$ imposes the
constraint $\hat Q=Q$. 
In so doing, we are really considering
a whole family of Anderson/Kondo models which includes 
under-screened models where $Q>K$ is larger than the number of
screening channels and the overscreened models where the number of
channels $K>Q$ exceeds the number of spinons per site.

We shall make the key assumption that the 
fully screened state where $Q=K$  develops an additional
stability with respect to the over and under screened states where
$Q\ne K$.
This assumption is motivated by the discovery of a spinon/holon gap in
the large $N$ Schwinger boson solution of the one and two impurity
Kondo models\cite{rech} and the 
leading Baym-Kadanoff approximation to the infinite $U$ Anderson
model\cite{lebanon}.  
The structure of these solutions lead us to believe that the gap in
the fractional particle excitation spectrum is robust in its
extension to finite $N$.   
The emergence of a gap in the infinite $N$ limit
provides a vital infra-red cuttoff to the fluctuations of the
fractional excitations. This cut off plays a dual role:
it stabilizes a Fermi liquid with a large Fermi surface, whilst at the
same time, providing a vital low-energy cut-off to the fluctuations of
the fractional holon and spinon fields removing them from the
low-energy Hilbert space. The presence of the gap guarantees that the
$1/N$ expansion contains no non-perturbative infra-red divergences,
and this in turn guarantees its survival over a finite range of $1/N$ values.

Suppose
$E_{0}$ is the energy of the ground-state with $Q=K$ 
 and $E_{\pm}$ be the ground-state energies of the state where
where $Q= K\pm 1$
is modified by one unit, then the ``commensuration gap'' associated
with complete screening of the moment at site $j_{0}$ is 
\begin{equation}\label{}
\Delta_{g}  = \frac{1}{2}[E_{+}+ E_{-}-   2 E_{0}].
\end{equation}
This gap is manifested as a gap in the
spectral functions of the spinon and the holon as illustrated in Fig
\ref{fig0}.
\figwidth=\columnwidth
\fg{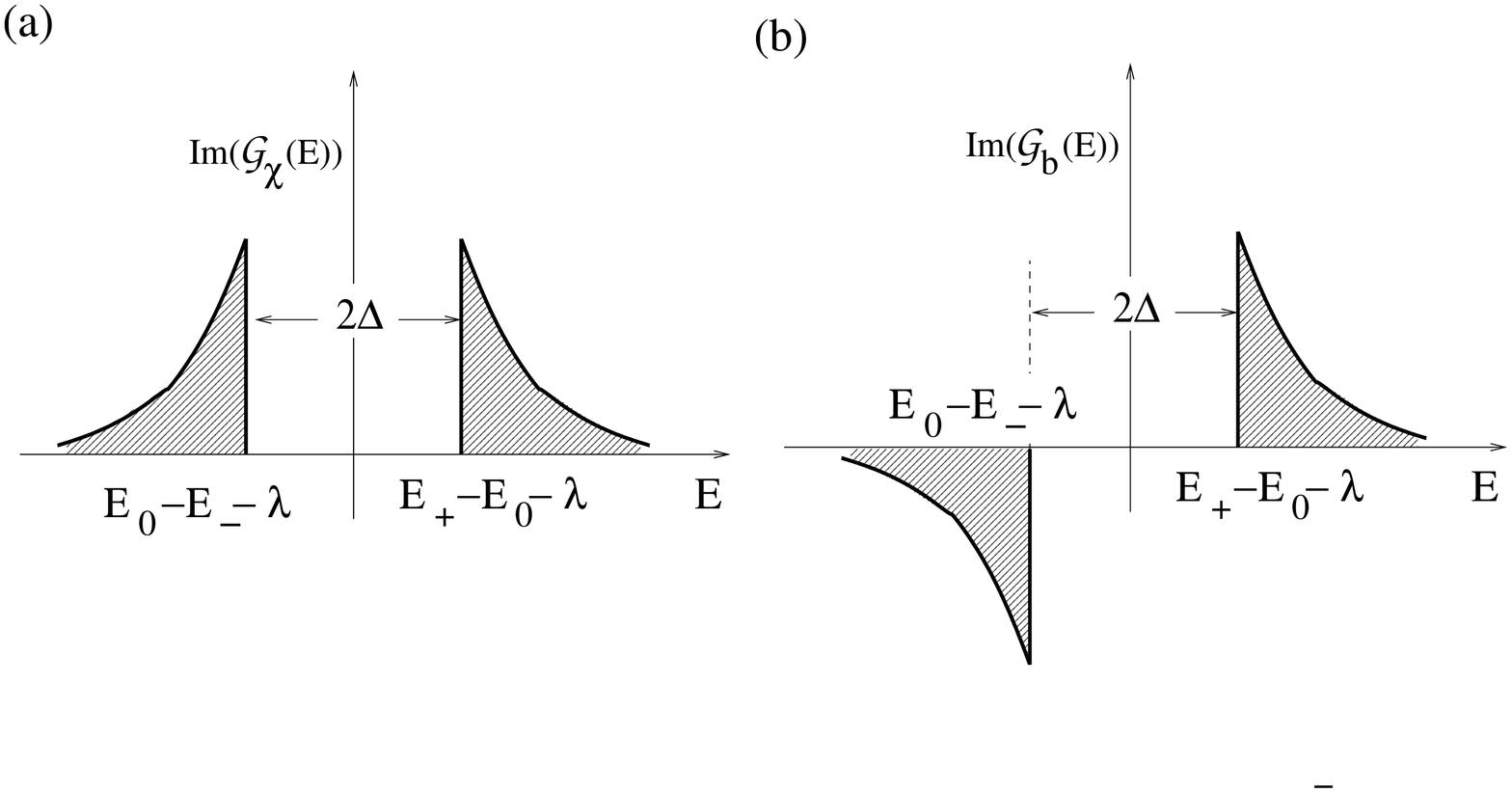}{fig0}{Illustrating the spectral gap in the holon and
spinon spectra. Creation of a holon or spinon 
increases $Q\rarrow K+1$, whereas the destruction of a holon
or spinon decreases $Q\rarrow K-1$.  }
With a gap we can calculate the low-temperature properties
of the model in the grand-canonical ensemble where remove the
constraint and associate a
chemical potential $\lambda$ with the conserved charge $Q$
\begin{equation}\label{}
Z_{G}[\lambda]\equiv e^{- \beta F[\lambda]} = {\rm  Tr} \left[e^{-\beta \bigl [H-
\lambda (Q-K)\bigr ]
} \right].
\end{equation}
Although the chemical potential $\lambda$ only imposes $\hat  Q=Q$ on
the average, the 
existence of  commensuration gap  guarantees that there is a finite
range of $\lambda$
\[
\lambda\in [ E_{0}- E_{-}, E_{+}- E_{0}],
\]
for which the constraint is imposed with exponential accuracy once 
$T \ll \Delta$.  

Our approach makes use of Coleman-Paul-Rech 
generalization of the Luttinger Ward Free energy functional,
to the gauge theory of interacting bosons and
fermions\cite{eliashberg,blaizot,rech} which can be compactly written as 
\begin{equation}\label{notsobigdeal}
F = T \Str \left[\ln  \left(- {\cal G}^{-1} \right)+
 \Sigma {\cal
G} \right]+ Y[{\cal G}],
\end{equation}
where 
 $\Str[A] = \Tr[A_{B}]-\Tr[A_{F}]$ denotes the supertrace of a
matrix containing both bosonic (B) and fermionic (F) components
(where the underline notation is used to denote a sum over internal
frequencies and a trace over the internal quantum numbers 
of the matrix). $\cg 
= (\cg _{0}^{-1}- \Sigma )^{-1}$ 
is the matrix describing the fully dressed Green's 
function of all elementary particles and fields entering the
Lagrangian, including the slave particles, where  $\Sigma $ is the
self-energy 
matrix and ${\cal G }_{0}$ the bare propagator of the fields.
Diagrammatically the propagators for the Anderson model are denoted as follows
\[
\centerline {\includegraphics[width=0.8\columnwidth]{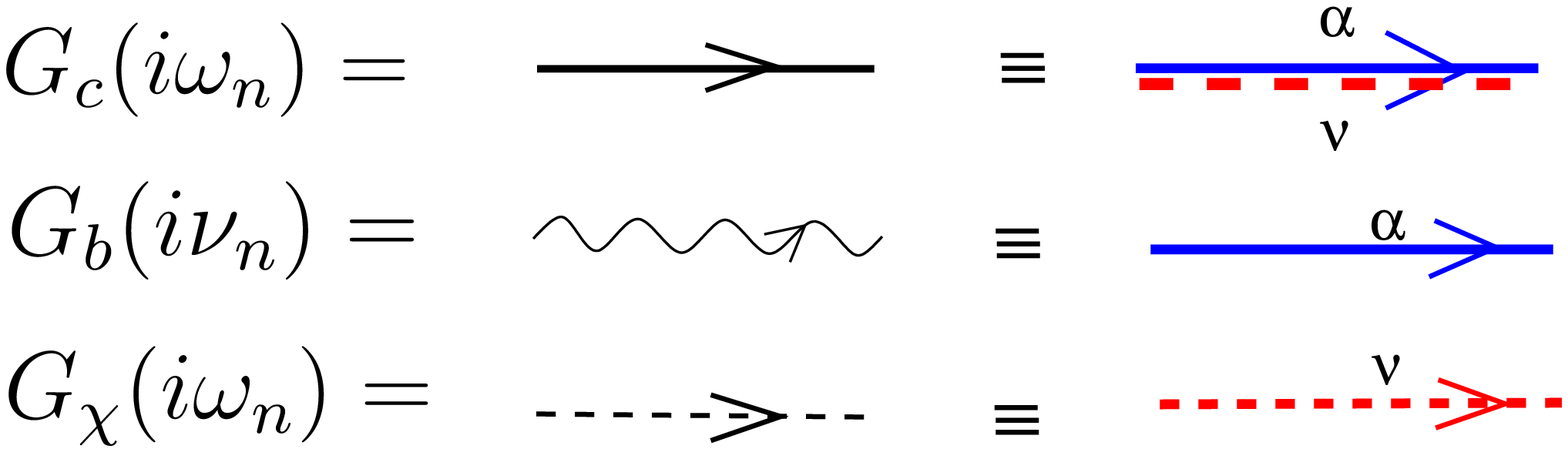}}\qquad .
\]
\noindent In the above expressions we display two alternative
notations for the propagators. In the first column, we adopt a
traditional  notation, using continuous, wavy and dashed lines to
represent the conduction, spinon and holon propagators
respectively. In the second column, we display the alternative
``railway
track'' notation that we use in cases where it is necessary to clearly
show the flow of spin and charge in the respective propagators.
At the interaction vertices
\[
\centerline {\includegraphics[width=2.in]{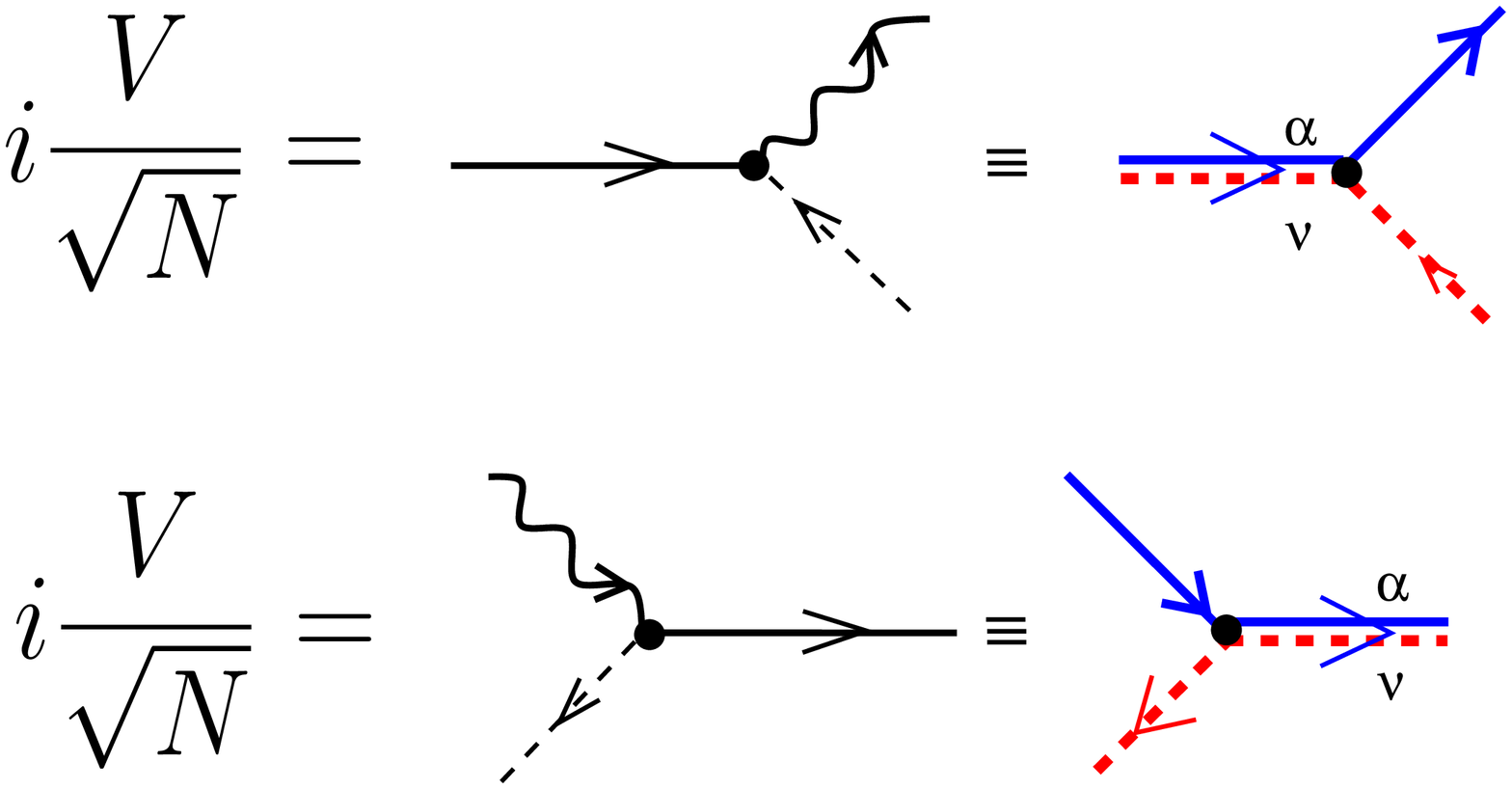}}\qquad ,
\]
\noindent the spin and charge of the electron fields divides up amongst the
holon and spinon fields. In the Feynman diagrams, each vertex is
associated with a factor $i V/\sqrt{N}$.
The generating functional 
$Y[\cg]$ may be written as the sum of all closed-loop two-particle
irreducible skeleton
Feynman diagrams. These can be ordered in a $1/N$ expansion as follows
\begin{equation}\label{genfunc}
\centerline {\includegraphics[width=2.5in]{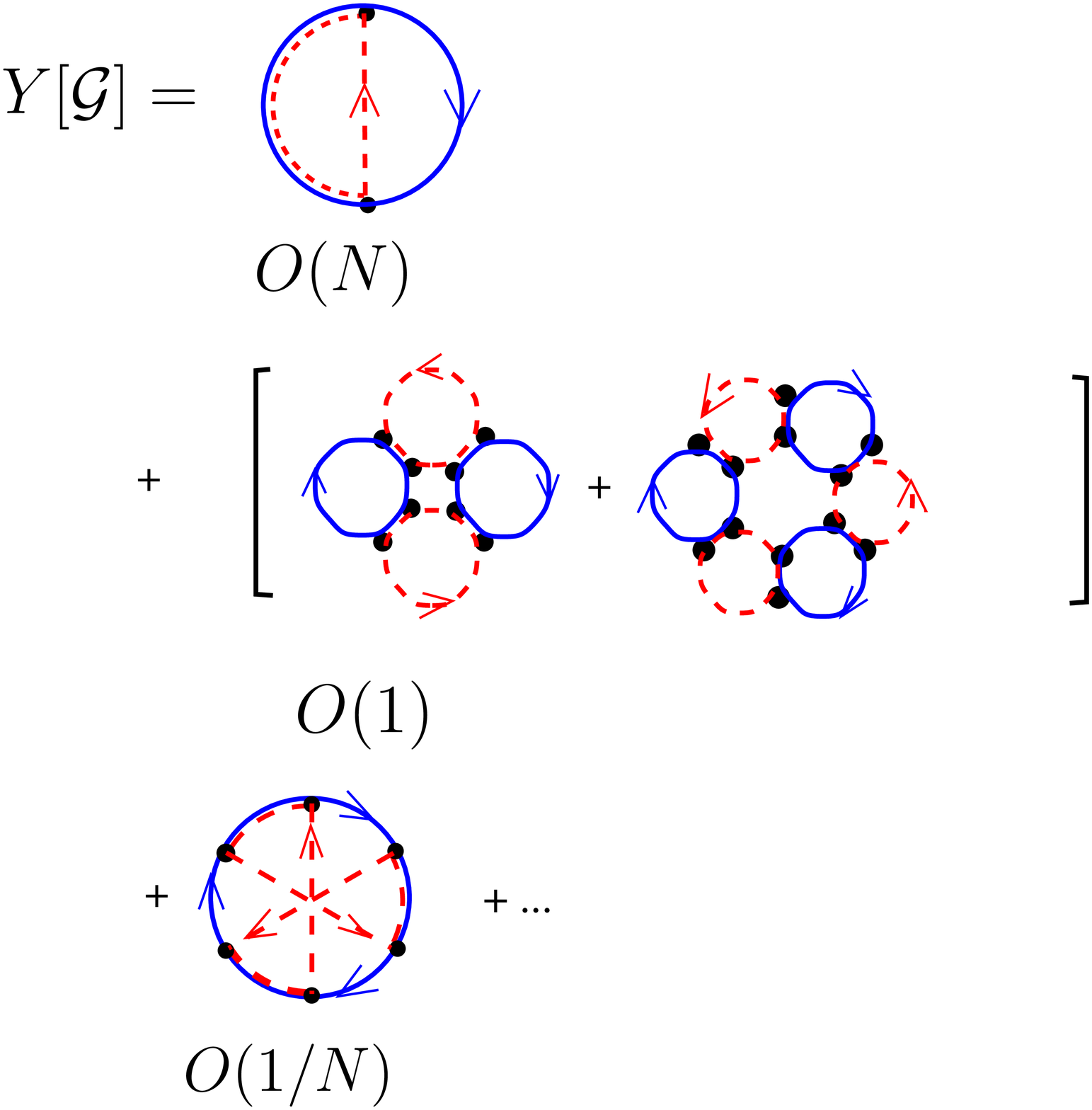}}\qquad .
\end{equation}
\noindent Each blue loop over spin gives a factor of $N$, whereas each red 
loop  over charge gives a factor of $K=k N$. 

The functional $Y[\cg ]$ is the
generating functional for the self-energies of the fields, so that 
variation of $Y[\cg ]$ with respect to the full Green's functions 
$\cg_{\zeta} $ of the conduction, spinon and holon 
fields, $\Sigma_{\zeta} =-\eta_{\zeta}\beta \ 
\delta Y / \delta { G}_{\zeta}$, 
self-consistently determines the self-energies $\Sigma_{\zeta}$ of 
these fields\cite{blaizot,indranil05}, where $\eta_{\zeta}
= \pm 1$ for bosons/fermions. 
We refer to  $Y[\cg ]$ as the
Luttinger Ward functional. Approximations to this functional are the
basis of Kadanoff-Baym conserving many body approaches\cite{kadanoffbaym}. 

The key difference between our new approach, and the classic Luttinger
Ward approach to the interacting electron gas, is the appearance of
the holon and spinon fields.  In the Luttinger Ward approach, the
$Y[\cg ]$ is a generating functional for the self-energies of the
fields.  The second-derivative of the generating functional with
respect to the Green-functions generates the off-shell scattering vertex
function between the fields $\Lambda = \beta ^2 
\delta^{2}Y[\cg]/\delta \cg^{2}$.
Nozi\`eres and Luttinger\cite{luttingernozieres} demonstrated that the multiple forward
scattering generated by this vertex gives rise to the on-shell Fermi
liquid amplitudes between the electron quasiparticles. The Landau amplitudes
are determined as solutions of the Dyson equation represented by the
Feynman diagrams,\\

\centerline {\includegraphics[width=3in]{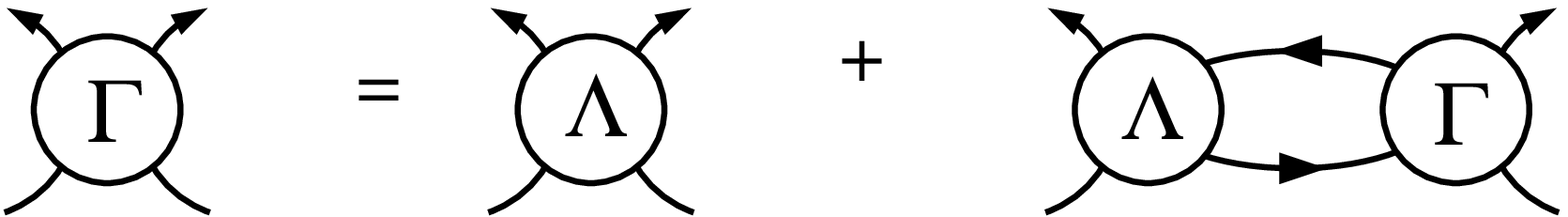}}.

\noindent This feature is preserved in the new approach, but  the
off-shell scattering vertex also involves 
virtual holon or spinon pairs, as shown below:
\\

\centerline {\includegraphics[width=3in]{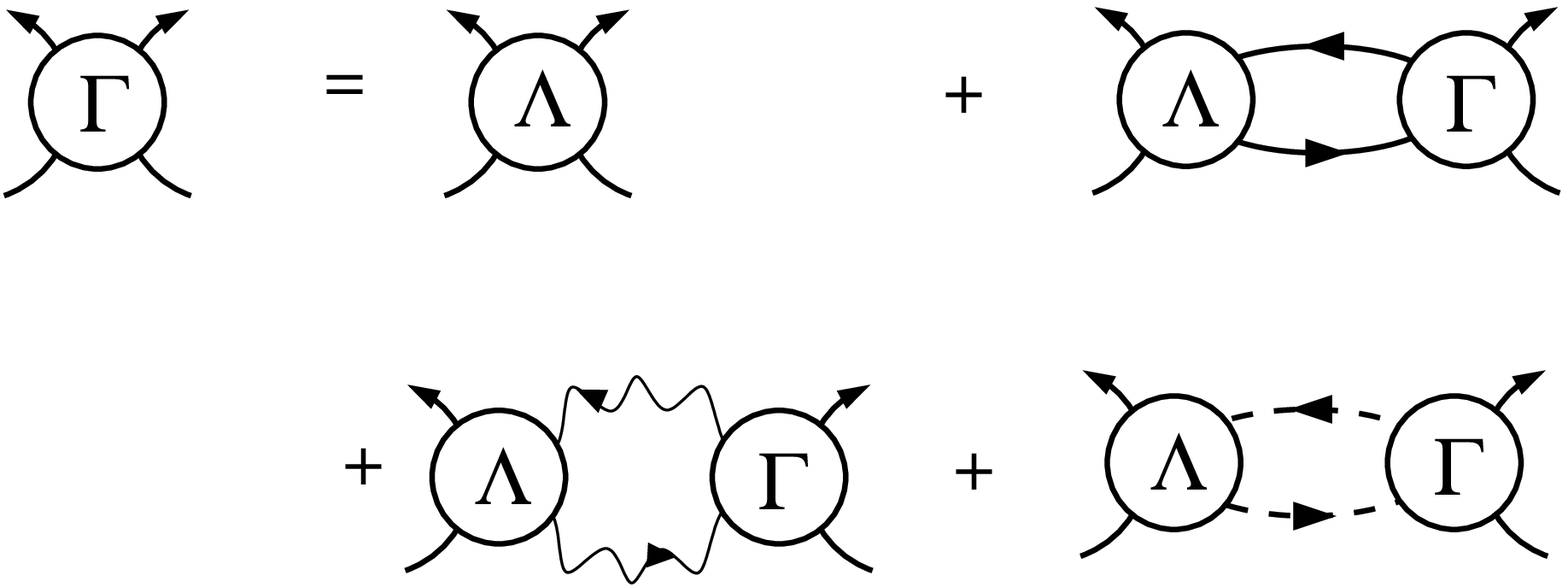}\label{fig2}.}

\noindent 
In this way, scattering off the virtual fractionalized excitations
determines the scattering amplitudes of low energy Fermi liquid.  
Remarkably, despite these effects,
all of the key conservation properties of the Landau Fermi liquid are
preserved, so long as the fractionalized excitations remain gapped. 

In the context of impurity models, the Landau Fermi liquid becomes
Nozi\`eres'  ``local'' Fermi liquid\cite{nozieres,blandin}, in which the 
energies of each electronic quasiparticles are expressed  in terms of the elastic
scattering phase shift $\delta_{\nu \alpha }$, where $\nu $ is the
channel index and $\alpha $ the spin index.   If $\Delta \epsilon
$ is
the spacing of states in the continuum, then the energy of each
electron in the continuum is shifted down by an amount $-
\left(\frac{\delta_{\nu \alpha \mu}}{\pi} \right)\Delta \epsilon$. The
Fermi liquid theory is then defined in terms of the dependence of the
phase shifts on the quasiparticle occupancies
\begin{equation}\label{}
\delta_{\nu \alpha }[\epsilon , \{n_{k\alpha \nu} \}] = \frac{\pi}{N}+
\epsilon \delta ' + \sum_{k',\nu',\alpha'}
\phi_{\nu \alpha , \nu'\alpha'} 
\delta n_{k\nu ' \alpha '},
\end{equation}
where the impurity Landau parameter
\begin{equation}\label{}
\phi_{\nu \alpha , \nu'\alpha'} 
= \langle \nu, \alpha ; \nu' \alpha
'\vert  \hat  \phi\vert \nu, \alpha ; \nu' \alpha\rangle 
\end{equation}
is the expectation value of the interaction energy between
quasiparticles of flavor/spin $(\nu,\alpha)$ 
and  $(\nu',\alpha')$.

By comparing the Free energy of the infinite $U$ Anderson model obtained in the
Luttinger Ward approach, we are able to diagrammatically identify each
of the terms in the above expansion. 
In particular, 
the conduction electron phase shift is given by
\[
\delta_{\nu \alpha } (\epsilon) = {\rm Im} \ln (1 - i \pi\rho \Sigma_{c}
(\epsilon-i0^+ )),
\]
where $\Sigma_{c} (\epsilon) $  is the conduction electron self-energy,
$\rho $ the density of conduction electron states at the Fermi
surface and $0^+$ is a small positive infinitesimal. We are also able 
to identify the Landau parameter with the on-shell forward scattering 
amplitude between quasiparticles
\[
\phi_{\nu \alpha , \nu'\alpha'} 
= \Gamma_{\nu \alpha , \nu'\alpha'} (\omega=0,\omega'=0).
\]
As in the finite $U$ Anderson model, each of these quantities enjoys expression entirely in
terms of conduction electron states.

The ability to link the quasiparticle physics with the gauge theory
diagrammatics
enables us to 
derive the key
conservation relationships for the 
infinite $U$ Anderson
impurity model, working to all orders in the $1/N$ expansion.
These relationships survive in the leading Kadanoff Baym\cite{lebanon}
approximation that defines the large $N$ limit, and we can illustrate
them explicitly in practical calculations. The key conservation laws
that appear from this approach are

\begin{itemize}

\item  The Friedel sum rule\cite{friedel,langer,langreth}.
\begin{equation}
\delta_c = \frac{\pi}{N} \frac {K -  n_{\chi} }{K} + O\left( 
\frac{T_K}{N D}\right).
\label{phase_shift}
\end{equation}

\item  The Langreth relation\cite{langreth} 
\begin{equation}\label{}
{\rm Im}t (\epsilon-i0^+)\vert_{\epsilon=0}= {\rm sin}^2 \delta_c/\pi\rho ,
\end{equation}
where $\rho $ is the density of states of the electron fluid.

\item  The Yamada-Yosida-Yoshimori relationship\cite{yamada,yoshimori} between the spin,
charge and channel susceptibilities and the linear specific heat coefficient
\begin{equation}
NK\frac{\gamma}{\gamma^0} = 
K\frac{N^2 -1}{N+K} \frac{\chi_s}{\chi_s^0} + 
N\frac{K^2-1}{K+N} \frac{\chi_f}{\chi_f^0} 
+\frac{\chi_c}{\chi_c^0}.
\label{WilsonRatio}
\end{equation}

\item The Korringa Shiba relationship\cite{shiba} between the dynamical spin
susceptibility and the impurity susceptibility
$\left. \chi''(\omega)/\omega\right|
_{\omega = 0}=  (N\pi/K) (\chi/N)^2$. 

\end{itemize}

The structure of the paper is as follows. In section (3)  we set up
various formal preliminaries.  In (4) we 
recapitulate  the relationship between conservation laws and Ward
identities, using them to rederive a general set of Friedel sum rules. 
In (5) we show how these can be used to derive the Langreth\cite{langreth} relation
for the scattering t-matrix and a general set of Friedel sum rules. 
In (6),  we introduce the off-shell interaction vertices amongst the
electrons and the fractional particles. Using
the Friedel relations we express the 
charge, spin and flavor susceptibilities in terms of these vertices. 
In (7) we use our gap hypothesis to derive an expression for the low temperature Free energy and
use this to relate the specific heat, with 
the spin, charge and flavor susceptibilities. 
In (8) we  identify the Nozi\` eres interaction parameters\cite{nozieres,blandin} 
with the
on-shell interaction vertices of the gapless electrons and use our
results to derive an identity between the linear specific heat
coefficient and a linear combination of the spin, charge and
flavor susceptibilities. Our result here is a generalization of the 
Yamada-Yosida-Yoshimori\cite{yamada,yoshimori} identities for the finite $U$ Anderson model. 
In section (9) we derive the Korringa-Shiba  relationship\cite{shiba} between the
uniform susceptibilities and the power-spectrum at low
energies. Finally in
(10) we discuss the extension of these ideas to a lattice
environment.

\section{Preliminaries.}\label{}

Our starting point is the infinite $U$ Anderson model.
\newcommand{\sk}{\vskip 1 truein \noindent 
}
%\sk
\begin{eqnarray}
{\cal H} &=& \sum_{{\vec k}\nu \alpha} \epsilon_{\vec k}
c^{\dagger}_{{\vec k}\nu \alpha } c_{{\vec k}\nu \alpha}+ 
\frac{{V}}{\sqrt{N}}\sum_{{\vec k}\nu \alpha}
\left[ c^{\dagger}_{{\vec k}\nu \alpha} \chi^{\dagger}_{\nu} 
b_{\alpha} +{\rm H.c} \right] \nonumber \\
\ &+& \epsilon_0 \sum_{\alpha} b^{\dagger}_{\alpha} b_{\alpha} 
-\lambda ( Q - 2S ),
\label{model}
\end{eqnarray}
where the spin components $\alpha$ are taken to run from $-j$ to
$+j$, where $N = 2j+1$, and the flavor components $\nu$ are taken 
to run from $-f$ to $+f$, where $K = 2f+1$.
The bare conduction electron, spinon and holon Green functions are
given by
\begin{eqnarray}\label{l}
\cg^{(0)}_{c \vec{k},\vec{k}'}(z) &=&
\frac{1}{z - \epsilon_{\vk }}\delta _{\vec{k},\vec{k}'} ,\cr
\cg^{(0)}_{b }(z) &=&
\frac{1}{z - (\epsilon_{0}-\lambda)} ,\cr
\cg^{(0)}_{\chi }(z) &=&
\frac{1}{z + \lambda} ,
\end{eqnarray}
where we use the symbol $z$ to denote the analytic extension of the
Matsubara frequencies into the complex plane. 
In an impurity model, it is generally convenient to trace the
conduction electron propagator over its momentum indices to obtain the
local propagator. The bare local conduction propagator is given by
\begin{equation}\label{}
\cg^{(0)}_{c } (z) =\sum_{\vk }
\frac{1}{ z  - \epsilon_{\vk }} \approx \mp  i \pi \rho \ \ \ \ {\rm for} \
z = \omega \pm i0^+ ,
\end{equation}
where the expression on the right is the large band width limit of the
local propagator, choosing the negative sign for the retarded
propagator ($z$ above the real axis) and the positive sign for the
advanced propagator ($z$ below the real axis).

When interactions are turned on, each of the fields acquires a
self-energy correction in the Full propagators
\begin{eqnarray}\label{l}
\cg_{c } (z)
 &=&[ [\cg^{(0)}_{c} (z)]^{-1}-\Sigma_{c} (z)]^{-1} ,
\cr
\cg_{b } (z)
 &=&[ z - (\epsilon_{0}-\lambda)
- \Sigma_{b} (z)]^{-1},
\cr
\cg_{\chi }  (z)&=&
[z  + \lambda 
- \Sigma_{\chi } (z)]^{-1}.
\end{eqnarray}

\begin{widetext}
Each of these self-energies is obtained by differentiating the
Luttinger Ward functional, shown in (\ref{genfunc})
with respect to the Green's functions
\bxwidth=1.7in
\upit=-0.3truein
\begin{eqnarray}\label{largeNself}
\cr
\Sigma_{b} (\omega) &=& -\beta \frac{\delta Y}{\delta \cg_{b} (\omega)}
= \  \  \frmup{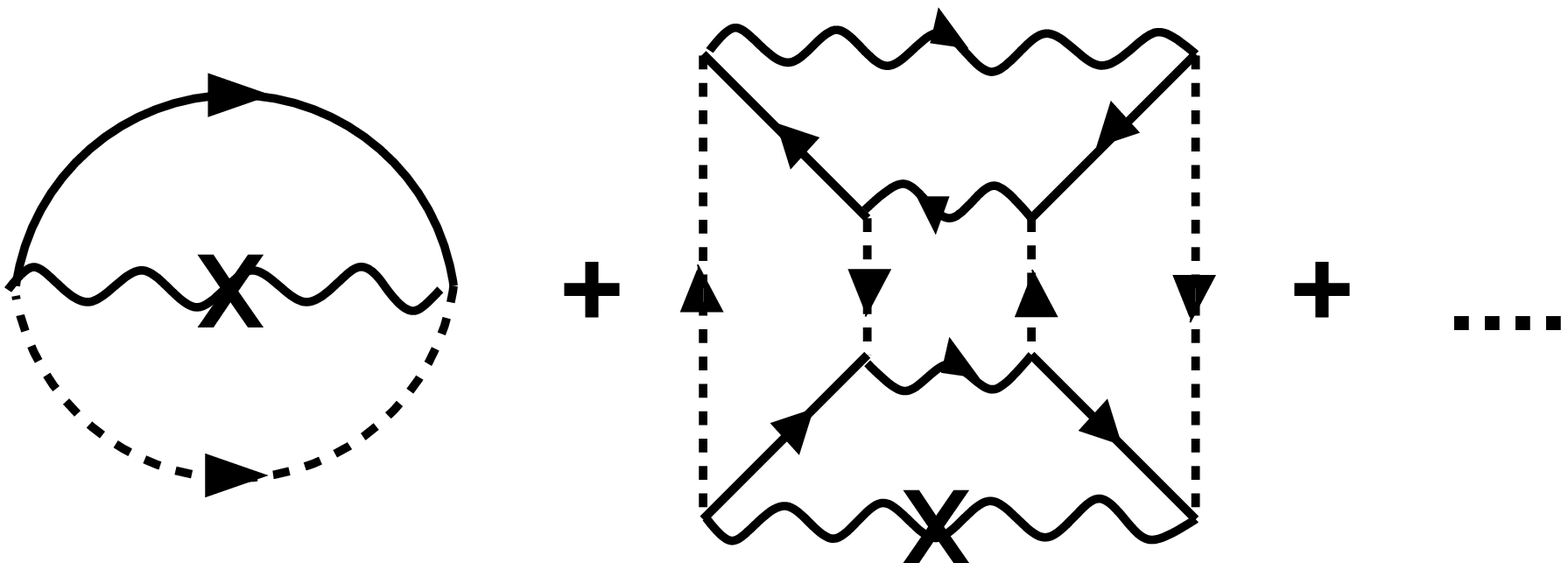} 
\ = \raiser{\epsfig{file=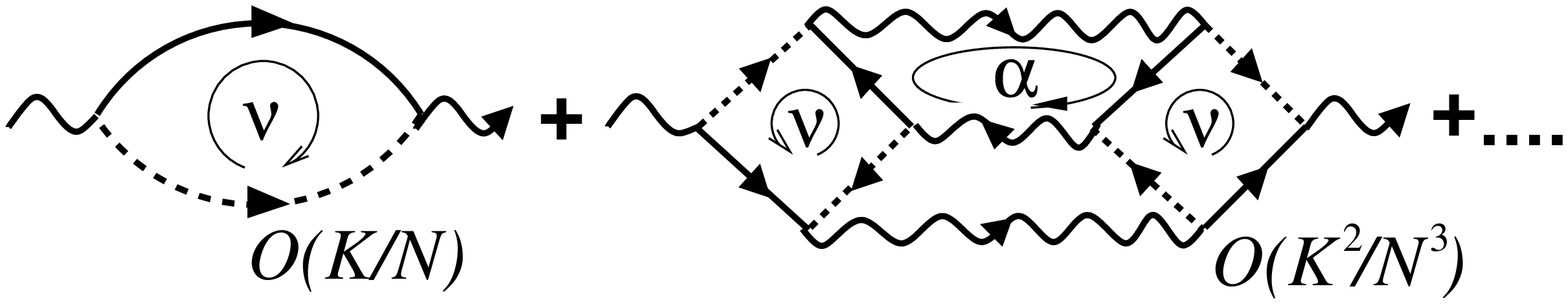,width=2.8in}} \ \ ,
\cr\cr\cr\cr
\Sigma_{\chi } (\omega)&=&+\beta \frac{\delta Y}{\delta \cg_{\chi }
(\omega)} 
= \ \  \frmup{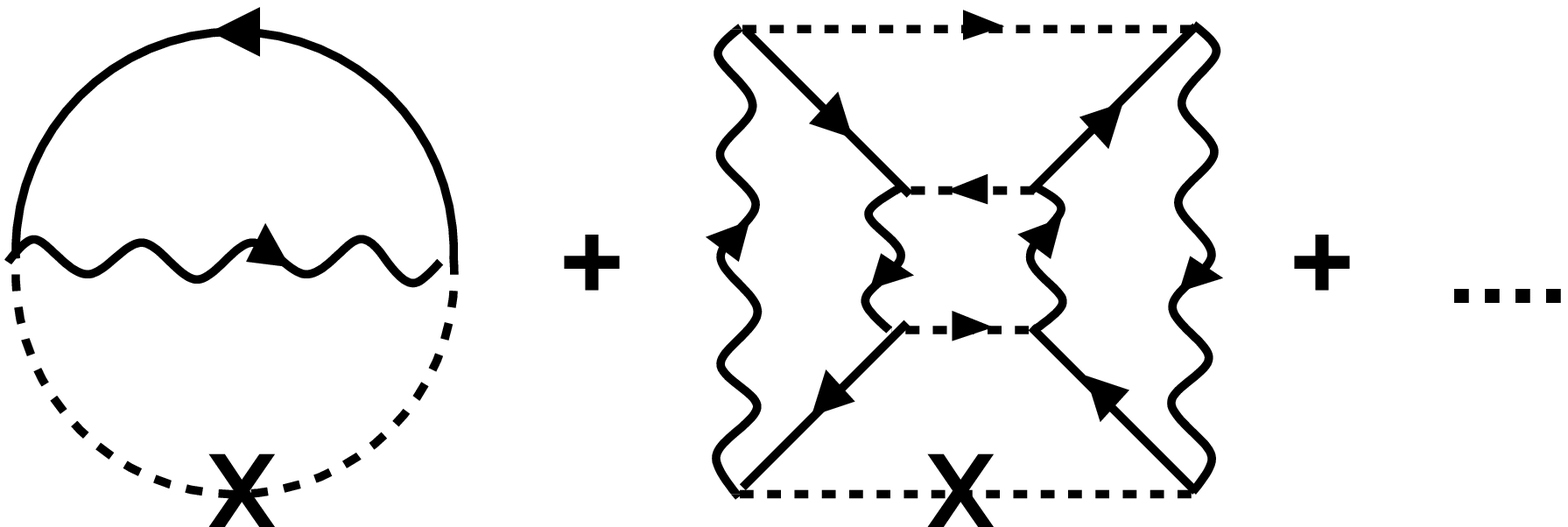}
\ = \raiser{\epsfig{file=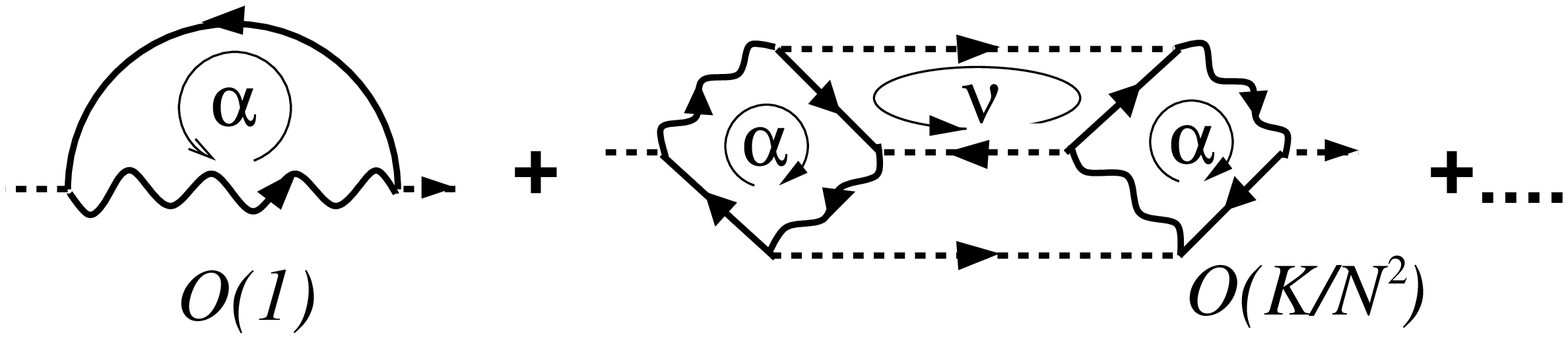,width=2.8in}} \ \ ,
\cr\cr\cr\cr
\Sigma_{c} (\omega)&=&+\beta \frac{\delta Y}{\delta \cg_{c} (\omega)}
= \ \  \frmup{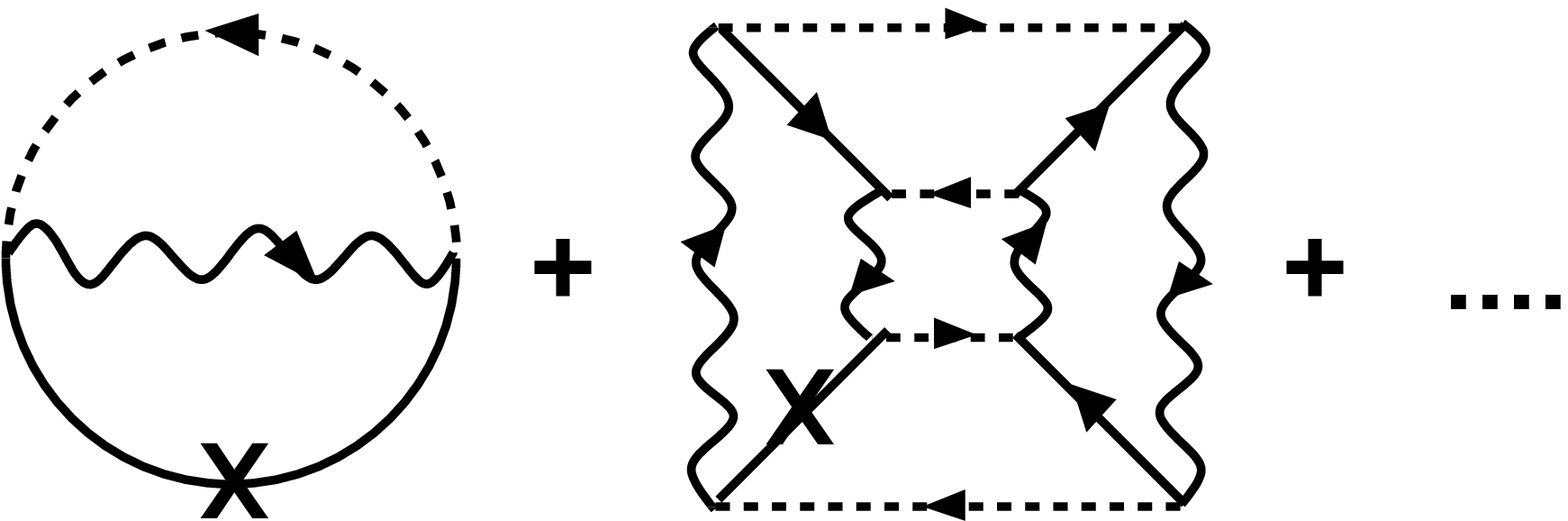} 
\ = \raiser{\epsfig{file=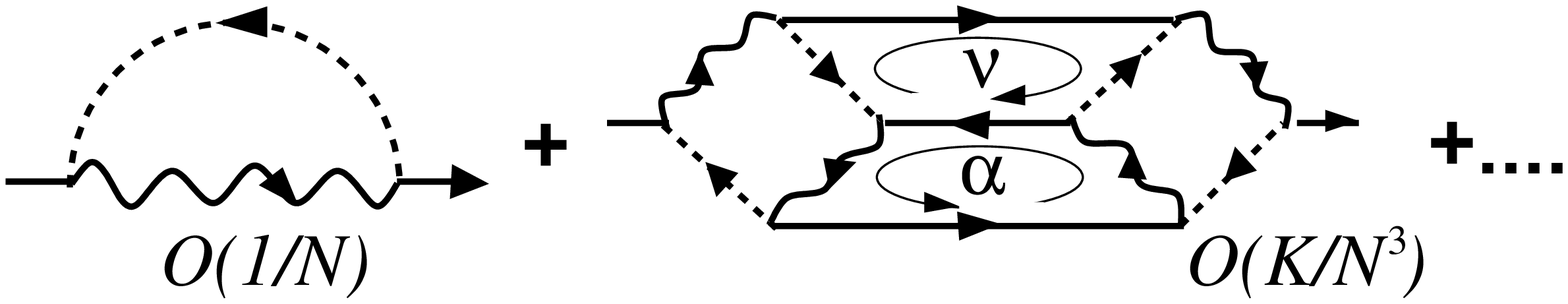,width=2.8in}} \ \ ,
\nonumber \\
\end{eqnarray}
\vskip 0.3in
\noindent where the cross indicates the line which is eliminated by
the functional differential. 
Although our discussion today
will focus on keeping all orders in this expansion, it is interesting
to briefly reflect on the leading terms in the diagrammatic expansion.
The leading order contributions to the self energies derive from the
virtual separation of the electron into spinon-holon pairs.  
These
terms are already sufficient to develop a conserving approximation
with all the key features of the infinite order resummation.  
The second-order terms in this expansion contain the sub-leading
effects of electron-electron scattering. 
\end{widetext}
\section{Friedel Sum rules and Luttinger Ward Identities.}\label{}

The Fermi liquid description of this model centers around the
relationship between the conduction electron phase shift and the 
thermodynamics, and the charge, spin and flavor susceptibilities.
We begin with a derivation of the Friedel sum rule. 
This part of the
paper closely follows earlier work on the Kondo model

In addition to the local conserved charge $Q= n_{b}+ n_{\chi }$, 
there are three globally conserved, physical 
quantities: charge ($Q^{C}$),
magnetization ($M=Q^{S}$) and flavor ($F=Q^{F}$)
associated with this model, which we write as 
\begin{eqnarray}\label{l}
{\hat Q}^C &=& \sum_{{\vec k}\nu \alpha}
 c^{\dagger}_{{\vec k}\nu \alpha } c_{{\vec k}\nu \alpha} 
- \sum_{\nu} \chi^{\dagger}_{\nu} \chi_{\nu},\cr
{\hat Q}^S &=& \sum_{{\vec k}\nu \alpha} \tilde{\alpha } \
c^{\dagger}_{{\vec k}\nu \alpha } c_{{\vec k}\nu \alpha}
+ \sum_{\alpha} \tilde{\alpha } b^{\dagger}_{\alpha} b_{\alpha} = \hat M ,
\cr
{\hat Q}^F &=& \sum_{{\vec k}\nu \alpha} \tilde{\nu } \ 
c^{\dagger}_{{\vec k}\nu \alpha } c_{{\vec k}\nu \alpha}
- \sum_{\nu} \tilde{\nu } \chi^{\dagger}_{\nu} \chi_{\nu},
\end{eqnarray}
respectively, 
where 
$\tilde{\alpha}= {\rm sign}(\alpha )$ is one for ``up
spins'' and minus one for ``down spins'', and 
$\tilde{\nu}= {\rm sign} (\nu)$ is one for flavors $\nu>0$
and minus one for flavor index $\nu<0$. 	
Each of these conserved quantities is associated with a corresponding
external field, which we introduce into the Hamiltonian by writing
\begin{eqnarray}\label{l}
H&=& {\cal  H} - \Delta \mu \hat Q^{C} - B \hat M - B^{F }\hat Q^{F},
\cr
&=&{\cal H}-\sum_{A =C,S,F} B^{A}\hat Q^{A},
\end{eqnarray}
where we introduce the notation $(B^{C},B^{S},B^{F})\equiv (\Delta
\mu, B, B^F)$.
We shall write each of these conserved quantities using the short-hand 
\begin{eqnarray}\label{l}
\hat  Q^{A}=  \psi^{\dagger} \hat q^A \psi ,
\end{eqnarray}
where $\psi\dg\equiv (c\dg_{\vec k},b\dg,\chi\dg  ) $ denotes the complete 
spinor of electron, spinon and holon fields. 
\begin{table}[t!]\label{tab1}
\begin{tabular}{||l|| c| c | c ||} \hline
  &\multicolumn{3}{|c||}{conserved quantity}\cr
\hline
\ \ 
particle \ \ & 
$\quad 
q^{C}\quad $ & $\quad q^{S}\quad $ &  $\quad q^{F}\quad  $  
\\ 
\hline\hline
  $\quad c$           & $1$ & $\tilde{\alpha}$ & $\tilde{\nu}$ \\ \hline 
  $\quad b$           & $0$ & $\tilde{\alpha}$ & $0$ \\ \hline
  $\quad \chi$        & $-1$& $0$              & $-\tilde{\nu}$ \\ \hline
\end{tabular}
\caption{The diagonal  components of the conserved charge, spin and flavor operators
$q^{C}$,  $q^{S}$ and $q^{F}$, resolved into their conduction $(c)$,
spinon $(b)$ and holon $(\chi )$ components.}
\end{table}
We can write the expectation value
of these operators in terms of the triplet of electron, spinon and
holon Green functions 
\begin{equation}
\cg = \left\{ \cg_{c{\vec k}{\vec k'}} ,\cg_b , \cg_{\chi} \right\},
\end{equation}
as
%\sk
\begin{eqnarray}\label{l}
\langle \hat Q^{A} \rangle &=&  {\rm Tr} \left\{ 
\psi^{\dagger} \hat q^A \psi \varrho \right\} = 
- T
{\underline{Str}}\left[\hat q^{A}{\cal G }\right],
\end{eqnarray}
where $\varrho$ is the thermal density matrix and the supertrace 
denotes a trace over each particle species (with a minus sign for fermions) 
and their quantum numbers, and the 
underline beneath the super trace denotes a summation over Matsubara frequencies, 
\begin{eqnarray}\label{l}
{\underline{Str}}\left[\hat q^{A}{\cal G }\right]
&=&\sum_{i\alpha_n} Str \left[\hat q^{A}{\cal G } (i \alpha_{n})\right]
= \sum_{i\alpha_n} \eta {\rm Tr} \left[\hat q^{A}{\cal G } (i \alpha_{n})
\right] \cr &=&
\sum_{i\nu_{n}}{\rm Tr}\left[q^{A}{\cal G}_{B} (i\nu_n) \right]
- \sum_{i\omega_{n}}{\rm Tr}\left[q^{A}{\cal G}_{F} (i\omega_{n}) \right].\nonumber
\end{eqnarray}

An alternative, and convenient way to formulate the conserved
quantities at absolute zero, is to replace  $q^{A}$ by its more
general form, $q^{A}\rightarrow  
q^{A}\gamma (\omega)
$ where 
\begin{equation}\label{}
\gamma (\omega)=\frac{\partial \cg_{0}^{-1} (\omega)
}{\partial\omega } ,
\end{equation}
so that 
\begin{equation}\label{}
\langle \hat Q^{A}\rangle 
= - T
{\underline{Str}}\left[
\hat q^{A}\gamma \cg
\right].
\end{equation}
This is a particularly useful device in the single impurity model where we 
work with the local conduction electron propagator. In this case the 
charges do not couple trivially in the local propagator and we
need to introduce frequency dependent vertices with the external
fields.  To see this, note that 
the bare local propagator becomes
\begin{eqnarray}\label{l}
\cg_c^0 (z,B) &=& \sum_{\bk} [ ( z - \epsilon_{\bk })
+  q^A_c B^{A}]^{-1}\cr
&=& \cg^0_c (z)
-[\cg^0_c (z)]^{2}
 ( B^{A}q^A_c \gamma (z)) +O (B^{2})\nonumber,
\end{eqnarray}
where 
\begin{eqnarray}
\gamma_{c} (z) = 
\frac{\partial \left[ \cg_c^{0} (z) \right]^{-1}}{\partial z}
= 
\frac{ \sum_{\vec k} \left( 
                       \frac{1}{z - \epsilon_{\bk } } 
                      \right)^2 }
      {\left[ \cg_c^{0} (z) \right]^2 } 
\end{eqnarray}
is identified as the frequency dependent vertex function.
The conduction electron vertex $\gamma_{c} (\omega)$
vanishes 
in the wide band limit $\gamma_c \rightarrow 0$. This is the basis of
the famous ``Anderson Clogston'' compensation theorem\cite{clogston61,compensationnote},
according to which the conduction band polarization of charge, spin or flavor
degrees of freedom  in the ground-state
\[
\langle Q^{A}_{c}\rangle = {\rm Tr} \left[ q^{A}_{c}\gamma
G_{c}\right] = O (T_{K}/D),
\]
where $D$ is the electron band-width and $T_{K}$ is the Kondo
temperature. 

Ward identities are properties of the zero temperature self-energies, 
Green's functions and scattering vertices that result from the the 
conservation laws. A key step in making the transition from finite, 
to zero temperature, is the replacement of Matsubara sums at finite 
temperature by 
continuous integrals along the imaginary axis at low temperature
\begin{eqnarray}
T\sum_{\alpha_{n}} A(i\alpha_n) &\stackrel{T \rightarrow 0} {= }&
\int _{-\infty }^{\infty }
\frac{d\alpha}{2\pi } A(i \alpha)\cr 
&\equiv& 
\int _{-i\infty }^{i\infty }
\frac{d\omega}{2\pi i} A(\omega) .
\label{replacement}
\end{eqnarray}  
Here $\alpha_{n}$ represents  the Bose Matsubara frequencies
$\nu_n=2n \pi k_{B}T$ along bosonic propagator lines, 
or the Fermion Matsubara frequencies $\omega_{n }=
(2n+1)\pi k_{B}T$ along fermion lines. Our ability to make this replacement 
presumes the existence of well-defined scales associated with the 
excitation spectrum. This is a situation that may not be satisfied 
at a quantum critical point, or an overscreened Kondo model where the 
spinon and holon propagators may exhibit $E/T$ scaling, with temperature 
as the relevant excitation scale. However, away from a quantum critical
point, the conduction electron lines are characterized by a well-defined 
Fermi energy and the existence  of a well-defined gap in the spinon and
holon spectrum guarantees that the continuous version of the Matsubara
sums is valid for these lines too.
Provided the replacement (\ref{replacement}) is allowed the zero temperature 
conserved quantities may be written in the form
%\sk
%\noindent 
\begin{equation}
\langle \hat Q^A \rangle = - \int_{-i \infty}^{i \infty} \frac{d \omega}
{2 \pi i} {\rm Str} \left\{ \hat q^A \gamma(\omega) {\cal G}(\omega) \right\}.
\label{Qcont}
\end{equation}

The starting point for the development of Ward identities is the Luttinger 
Ward functional. The conserved charges are conserved at each vertex in the 
diagrams contributing to the generating function $Y$.   
For the quantity  $Q^{A}$, the charge
$q^{A}_{\zeta }$ is associated with each propagator line ${\cal G}_{\zeta}$ inside 
the functional. When we shift the frequency of each propagator by a 
constant differential times the corresponding charge 
\begin{eqnarray}
{\cal  G}_{\zeta}(\omega) \rightarrow {\cal G}_{\zeta} (\omega + \delta \omega^A_{\zeta}) ,
\ \ \ \ \ \ 
\delta \omega^A_{\zeta } = \delta \omega \  q^A_{\zeta },
\end{eqnarray}
the generating functional is unchanged. Since the derivative of the
generating functional with respect to the Green's functions is the
self-energy, this leads to the Ward identity
\begin{eqnarray}
\frac{\delta Y[{\cal G}]}{\delta \omega^A} = \int_{-i\infty}^{i\infty}
\frac{d \omega}{2 \pi i} {\rm Str} \left\{  \Sigma \frac{d {\cal G}}{d\omega} 
\hat q^A \right\} =  0,
\end{eqnarray}
and integration by parts then yields
\begin{equation}
\int_{-i\infty}^{i\infty} 
\frac{d \omega}{2 \pi i} {\rm Str} \left\{ \hat q^A \frac{d\Sigma}{d\omega}
{\cal G} \right\} =0.
\label{Ward}
\end{equation}
Equations (\ref{Qcont}) and (\ref{Ward}) sum up to an integral over a 
full differential, rotating the integration contour around the negative 
real axis we get the Friedel sum rules
\begin{equation}
\langle \hat Q^A \rangle =
 - \frac{1}{\pi} \ {\rm Im} \ {\rm Str} 
\left\{ \hat q^A \ln \left[ -{\cal G}^{-1}(-i0^+) \right] \right\}.
\end{equation}
In impurity problems, we are interested in change  in these quantities
that results from coupling the impurity to its environment, which is
given
by 
\begin{eqnarray}\label{l}
\Delta Q^{A}&=& \langle \hat Q^A \rangle -
\langle \hat Q^A \rangle_{0}
\cr
&=& - \frac{1}{\pi} \ {\rm Im} \ {\rm Str} 
\left\{ \hat q^A \ln \left[ {\cal G}^{-1}/ {\cal  G}_{0}^{-1}\right] 
\right\}\vert_{\omega= - i0^+ }\cr
\cr
&=& \left. 
 -\frac{1}{\pi}
 \ {\rm Im} \ {\rm Str} 
\left\{ \hat q^A \ln \left[ 1- {\cal  G}_{0} (\omega) \Sigma (\omega)
\right] 
\right\}\right\vert_{\omega= - i 0^+ }.
\end{eqnarray}
The quantities on the right hand side are the ``phase shifts''
of the electrons, spinons and holons. If we identify
\begin{equation}\label{}
\delta_{\zeta} = \left. {\rm Im} 
\left\{ \ln \left[ 1- {\cal  G}_{0\zeta} (\omega) \Sigma_{\zeta} (\omega)
\right] 
\right\}\right\vert_{\omega= - i 0^+ } ,\qquad (\zeta = c,b,\chi )
\end{equation}
then we can write
\[
\Delta Q^{A}= \sum q^{A}_{c\ \nu\alpha
}\frac{\delta^{c}_{\nu\alpha}}{\pi}
+ \sum q^{A}_{\chi \ \nu}\frac{\delta^{\chi }_{\nu}}{\pi}-\sum q^{A}_{b\ \alpha
}\frac{\delta^{b}_{\nu\alpha}}{\pi}.
\]
At first sight, this is very different from the expression we would
expect for a Fermi liquid, where there are no holon or spinon
contributions. However, 
so long as the spinons and holons are gapped, 
their corresponding phase shifts are either $\pi$ or $0$ and do not
vary with the external fields.
General arguments lead us to believe that in the impurity model, the
spinon phase shifts are zero, while the $\chi $ phase shifts are $\pi$.
To see this, consider the Friedel sum rule for the 
conserved charge $Q = n_{b}+n_{\chi }$, for which
$(q_{c},q_{b},q_{\chi })= (0,1,1)$. In this case, 
\begin{equation}\label{}
\Delta Q = 2S =K= -N \frac{\delta_{b}}{\pi} + K \frac{\delta_{\chi }}{\pi}.
\end{equation}
Provided all the
phase shifts are positive definite, then this quantity can never
exceed $K$ and the maximum value is only attained if $\delta_{b}=0$
and $\delta_{\chi }= \pi$. 
If the spinon
phase shifts are zero, then for a perfectly screened impurity,
with $K=2S$ it follows that the holon
phase shifts equal $\pi$ and the spinon phase shifts vanish so long as
the spinon-holon gap is preserved.

The general sum rules then become
\begin{eqnarray}\label{friedelsr}
\Delta Q^{C} &=& 
\sum_{\nu \alpha }\frac{\delta_{c\nu\alpha }}{\pi} 
- \sum_{\nu}\frac{\delta_{\chi  \nu }}{\pi}= 
\sum_{\nu \alpha }\frac{\delta_{c\nu\alpha }}{\pi} 
- K,
\cr
\Delta Q^{A} &=& \sum_{\nu \alpha }q^{A}_{c\nu\alpha}
\frac{\delta_{c\nu\alpha }}{\pi} 
\qquad (A=S,F),
\end{eqnarray}
where  the conduction electron phase shift is
\begin{eqnarray}\label{phaseshift}
\delta_{c \nu\alpha} = {\rm Im} \ln [1 - 
\cg_c^0 \Sigma_{c\nu\alpha }  ] \vert_{\omega=  - i 0^+ }.
\end{eqnarray}
The $\chi $ fermion contribution only enters into the charge Friedel
sum rule, 
since $\sum_{\nu }q^{C}_{\chi \nu }=0$ and $\sum_{\nu }q^{S,F}_{\chi \nu }=0$.
The first sum rule expresses the conventional Friedel  sum rule. 
In general, the impurity charge $\Delta Q^{C}$ contains a conduction
electron and a holon contribution
\begin{equation}\label{l}
\Delta Q^{C}= \Delta Q^{C}_{e} - \langle n_{\chi}\rangle .
\end{equation}
According to the Anderson Clogston theorem\cite{clogston61,compensationnote}, 
the first term is of order $T_{K}/D$, where $D$ is the band-width, and vanishes
in the infinite band-width limit. The remaining term is simply the charge
associated with the empty $f-$ states at the impurity, $\Delta Q^{C}=
-\langle n_{\chi }\rangle $.  Using the phase shift expression for  $\Delta Q^{C}$ in 
eq. (\ref{friedelsr} ), the Friedel
sum rule can then be rewritten as
\begin{equation}\label{}
\sum_{\nu \alpha }\frac{\delta_{c\nu\alpha }}{\pi} = K - \langle n_{\chi}\rangle.\end{equation}
In the case where there is no magnetic or flavor polarization, all
phase shifts are equal, $\delta_{c\nu\alpha}=\delta_{c}$ and
$\sum_{\nu\alpha }\delta_{c\nu\alpha}=NK \delta_{c}$, so that the
Friedel sum rule becomes
\begin{equation}\label{}
\delta_{c}=\frac{\pi}{N}\left(1 - \frac{\langle n_{\chi}\rangle}{K} \right).
\end{equation}

\section{Langreth Sum rule and the Kondo resonance}\label{}

The emergence of the Kondo resonance is one of the signature features
of the Kondo effect. 
The  ``Langreth sum rule''\cite{langreth}, links the Friedel sum rule
to the formation of the Kondo resonance by providing a rigorous
constraint on the size of the spectral function associated with the
localized state.

The t-matrix of the conduction electrons is determined by the spectral
function of the electrons localized in the magnetic moment,
\begin{equation}\label{}
t (i\omega_{n})= -\frac{V^{2}}{N}
\int_{0}^{\beta}\langle T X_{\alpha \nu}\dg  (\tau)X_{\alpha \nu} (0)\rangle e^{i\omega_{n}\tau}d\tau 
\end{equation}
We shall assume here that there is no magnetic or flavor polarization,
so that the t-matrix is identical for all $\nu $ and $\alpha $. 
The t-matrix is determined diagrammatically, by the process of
repeated scattering off the localized state, and is given by
\begin{equation}\label{}
\epsfig{file=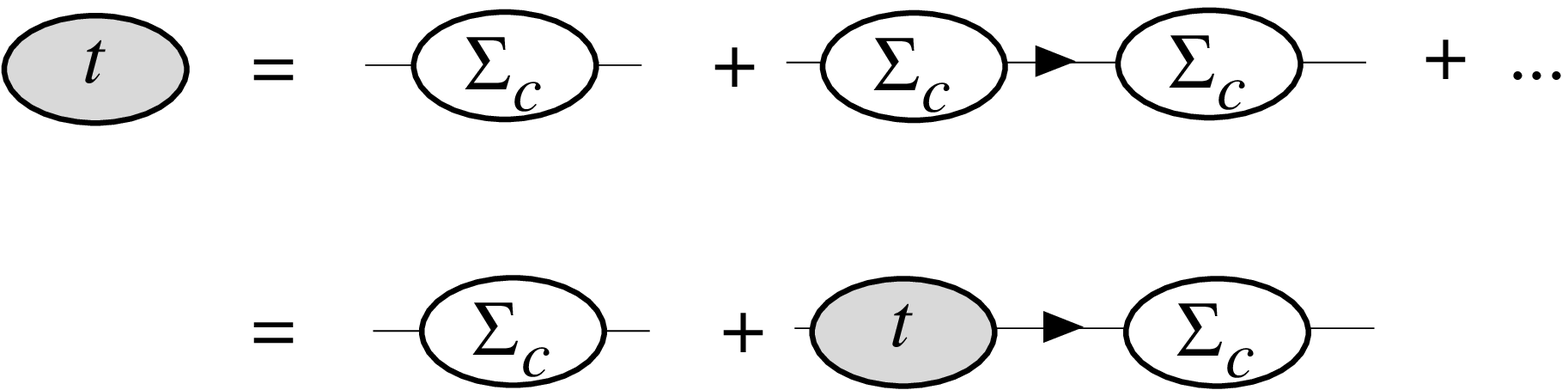,width=2.7in},
\end{equation}
\vskip 0.1in
or
\begin{eqnarray}\label{l}
t (z)&=& \Sigma_{c} (z)+ \Sigma_{c} (z)G_{c}^{(0)}(z)\Sigma_{c} (z)+ \dots
\cr
&=& \frac{\Sigma_{c} (z)}{1- G_{c}^{0} (z)\Sigma_{c} (z)}
\end{eqnarray}
From (\ref{phaseshift}) we see that denominator in the t-matrix is
also the argument of the scattering phase shift.  In the limit of a
broad band-width, we can replace $G_{c}^{(0)} (\omega+i\delta )= -i
\pi\rho $, so that the phase shift and t-matrix become
\begin{eqnarray}\label{relationships}
t (\omega+i\epsilon)&=& \frac{\Sigma_{c} (\omega+i\epsilon)}{1+i \pi\rho \Sigma_{c} (\omega+i\epsilon)}\cr
\delta^{c } &=& {\rm Im} \ln [1 + i \pi 
\rho\Sigma_{c}  ] \vert_{\omega=  - i 0^+ }.
\end{eqnarray}
If we examine the decay processes associated with the conduction
electron self-energy, we see there are two types of process. In one
type, the electron decays into holon, spinon combinations, but since
these are gapped excitations, they produce no contribution to the
inelastic electron decay processes at the Fermi energy. The only
remaining decay processes involve the production of electron-hole
pairs, such as
\bxwidth=1.7in
\upit=-0.25truein
\vskip 0.1in
\begin{eqnarray}\label{l}
\Sigma''_{c} (\omega)&\sim &Im \left[ \phantom{\int_{\int_{a}}^{B}}\hskip -0.2in
\raiser{\epsfig{file=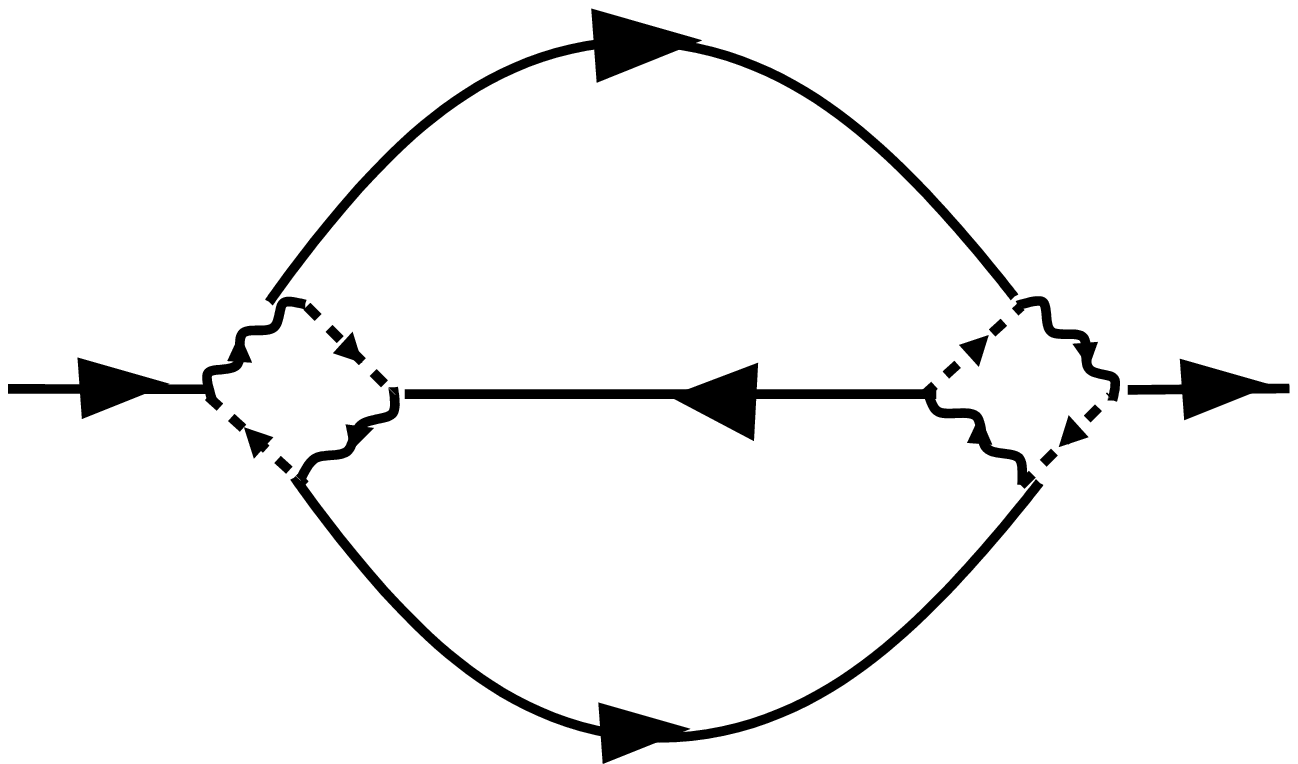,width=1.0in}}\ \right]\cr\cr\cr
&=&
\pi\sum_{k,\ k',\ k''} \left| \phantom{\int_{\int_{a}}^{B}}\hskip -0.2in
\raiser{\epsfig{file=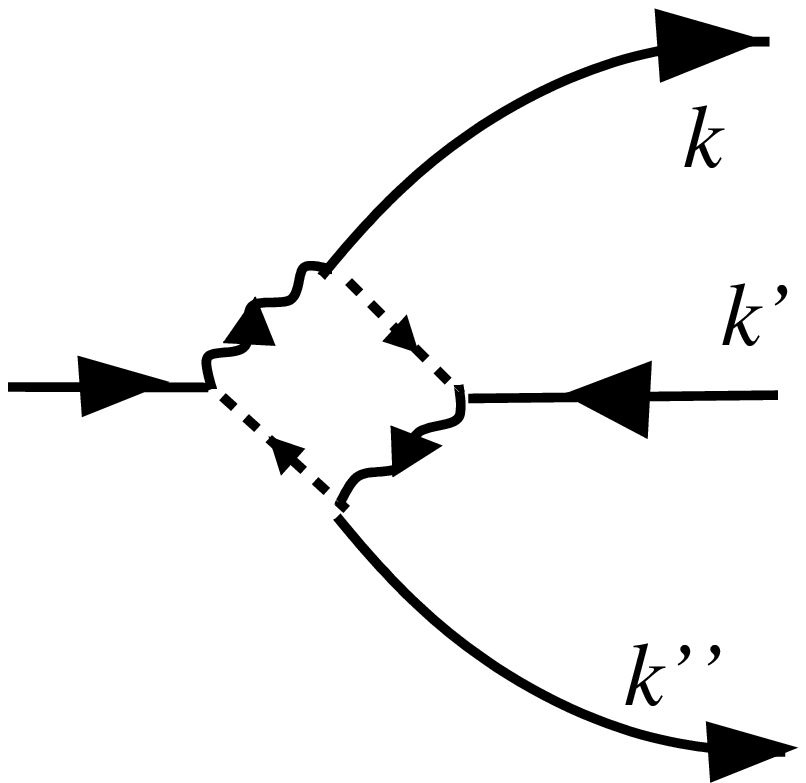,width=0.55in}}\
\right|^{2}f_{k'} (1-f_{k}) (1-f_{k''})\cr\cr\cr
&&\qquad \qquad \qquad \qquad 
\end{eqnarray}
\vskip 0.1in
\noindent 
where the internal loop of gapped fractional quasiparticles does not
contribute to the low-energy decay processes. 
These are the well-known decay processes of the conventional Fermi
liquid, producing an inelastic decay rate that grows quadratically
with energy and temperature $\Sigma''_{c} (\omega)\sim
[\omega^{2}+(\pi T)^{2}]$. At zero temperature and frequency, the
conduction electron self-energy is thus entirely real, $\Sigma_{c}''
(0)=0$. So from (\ref{relationships}),  $\tan\delta_{c}= \pi\rho \Sigma_{c} (0) $,
and the on-shell t-matrix is given by 
\begin{equation}\label{}
t (0+i\epsilon ) = \frac{1}{\pi\rho } \frac{\tan \delta_{c}}{(1 + i
\tan \delta_{c})}= \frac{\sin \delta_{c}}{\pi\rho }e^{-i\pi\delta_{c}},
\end{equation}
and the imaginary part of the t-matrix is then 
\begin{equation}\label{}
{\rm Im }t (0+i\epsilon) = \frac{\sin^2 \delta_{c}}{\pi \rho }.
\end{equation}
The existence of this quantity as an adiabatic invariant is well-known
in the finite $U$ Anderson model, and its appearance in the limit that
$U\rightarrow \infty $ is an important test of these methods. 

\section{Static Susceptibilities}\label{}

The susceptibilities of the impurity contribution to the conserved 
charges can now be obtained by differentiating the above expressions
with respect to the corresponding 
fields $\chi^A  = d   \langle \hat Q^A \rangle_{\rm imp} / dB^A $ at 
zero field $B^A =0$. Since the spinon and holon are gapped, their phase 
shift is pinned to the constant values of $0$ and $\pi$, and therefore 
the only contribution to the susceptibilities comes from derivatives of 
the conduction electron phase shift $\delta = {\rm Im} \ln [1 - 
G_c^0\Sigma_c ]$ :
\begin{eqnarray}
\chi_A  
= \left. \frac{1}{\pi} \sum_{\alpha \nu}  q^A_{c\alpha \nu}
\frac{d \delta_{ \alpha \nu}}{d B^A} \right|_{B^A = 0}. 
\label{susfromphase}
\end{eqnarray}  
Explicitly the derivatives of the phase shift are given by
\begin{equation}
\frac{d  \delta}{dB^A} = \left.  - {\rm Im}\cg_c 
\frac{d \Sigma_{c}}{dB^A} 
- {\rm Im} \left( 
\cg_c + \cg_c^0 \right) \gamma_c \hat q^A_c \right|_{-i0^+}\ ,
\end{equation}
and in the wide bandwidth limit in where $\gamma_{c}$ vanishes 
\begin{equation}
\frac{d  \delta}{dB^A} = \left.  - {\rm Im}\cg_c \    
\frac{d \Sigma_{c}}{dB^A} 
\right|_{-i0^+}.
\label{phasederivative}
\end{equation}

\begin{widetext}
To calculate the derivatives of the self energies it is useful to 
introduce the bare and full off-shell interaction vertices;
The bare interaction vertex $\Lambda$ is defined as the functional 
Hesseyan of the LW functional 
\begin{eqnarray}
\Lambda_{a a'} (\omega , \omega') = (2\pi i)^2 \eta_{a} 
\eta_{a'}\frac {\delta^2 Y[{\cal G}]}
{\delta \cg_a(\omega) \delta \cg_{a'}(\omega')} = 
%\cr \cr = 
2\pi i \eta_{a'} 
\frac{\delta \Sigma_a(\omega)} {\delta \cg_{a'}(\omega')}\ , 
\end{eqnarray} 
where $a$ and $a'$ list the particle species $c$, $b$ and $\chi$ and 
their corresponding indices $\alpha$ and/or $\nu$, while
$(\eta_{b},\eta_{c},\eta_{\chi })= (1,-1,-1)$ grade the anticommuting 
character of the Fermi fields. The $2\pi i$ factors come to compensate 
for the integration measure and replace the $1/T$ factors in the zero 
temperature limit when the discrete Matsubara frequencies are replaced by 
the continuous imaginary variable $\omega$. For example the interaction 
vertex between an electron and a spinon is 
\bxwidth=1.7in
\upit=-0.3truein
\begin{eqnarray}\label{Lambda_fig}
\cr
\Lambda_{cb}(\omega,\omega') = \frac {- (2\pi i)^2 \delta^2}{\delta \cg_c(\omega) 
\delta \cg_b(\omega')}  
\raiser{\epsfig{file=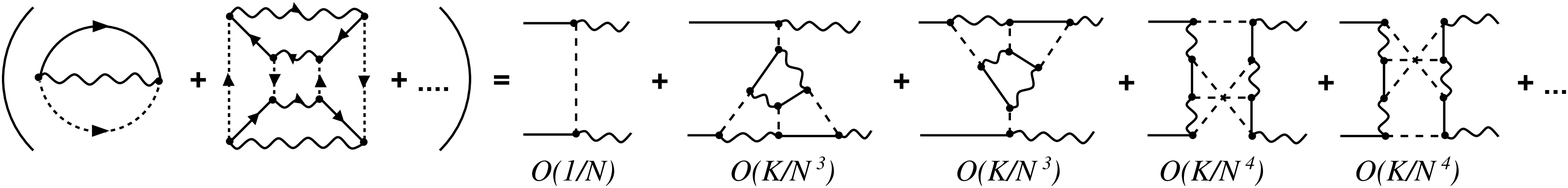,width=5.4in}}.  
\nonumber 
\end{eqnarray}
\vskip 0.3in
The derivatives of the self energies with respect to the fields $B^A$ 
are expressed in terms of the bare interaction vertex by
\begin{eqnarray}
\frac{d \Sigma_a (\omega) }{d B^A}
 = - \sum_{a'} \int 
\frac{d \omega'}{2 \pi i} \Lambda_{aa'}(\omega,\omega')\eta_{a'}
\cg^2_{a'}(\omega') 
% \times  \ \ \ \ \ \ \ \ \nonumber \\ 
\left\{ \gamma_{a'}(\omega') \ \hat q^A_{a'} - \frac{d \Sigma_{a'}
(\omega')}{dB^A} \right\}\ . \ 
\end{eqnarray}
\end{widetext}
adopting a heuristic matrix notation we may write this
\begin{equation}\label{l}
\frac{d \Sigma }{d B^A} = - \Lambda \eta \cg^{2}
\left(\gamma q^{A}- \frac{d \Sigma }{d B^A} \right),
\end{equation}
which implies that
\begin{equation}\label{l}
\left(1 - \Lambda \eta \cg^{2}\right)\frac{d \Sigma }{d B^A} = 
-\Lambda \eta \cg^{2}\left(\gamma q^{A}\right),
\end{equation}
or
\begin{equation}\label{l}
\frac{d \Sigma }{d B^A} = -\Gamma\eta \cg^{2}
\left(\gamma q^{A}\right),
\end{equation}
where
\begin{equation}\label{l}
\Gamma = \left(1 - \Lambda \eta
\cg^{2}\right)^{-1}\Lambda 
\end{equation}
is the solution to the Dyson equation
\[
\Gamma = \Lambda + \Lambda \eta \cg^{2}\Gamma.
\]
Restoring the indices, we then have
\begin{eqnarray}
\Gamma_{a a'}(\omega,\omega') = \sum_{a''} \int \frac{d \omega''}{2\pi i}
\Lambda_{a a''}(\omega,\omega'') \times 
\ \ \ \ \ \ \ \ \ \ \ \ \ \ \ \ \ \nonumber 
\\ \left\{ 2\pi i\ \delta_{a'',a'} \ 
\delta (\omega''-\omega') + \eta_{a''} \cg^2_{a''}(\omega'') \ \Gamma_{a'' a'}
(\omega'',\omega') \right\} .\nonumber
\end{eqnarray}
It is instructive to represent this equation diagrammatically. For
example, the vertex with incoming and outgoing electrons is
$\Gamma_{cc}$, represented as
\vskip 0.1in
\centerline {\includegraphics[width=2.8in]{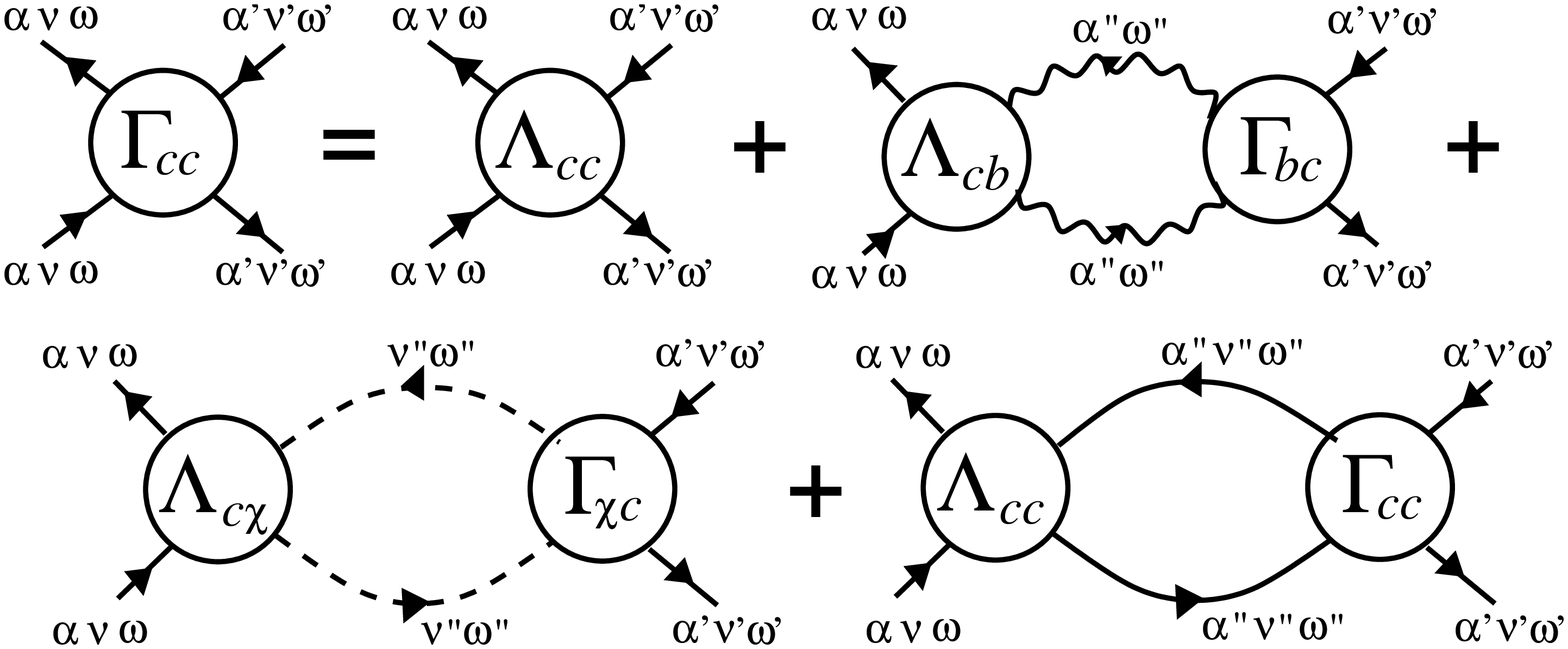}\qquad
.\label{Gamma}}
\vskip 0.1in
\noindent The self energy derivatives are then 
reduced to 
\begin{equation}
\frac{d \Sigma_a (\omega) }{d B^A} = - \sum_{a'} \int 
\frac{d \omega'}{2 \pi i} \Gamma_{aa'}(\omega,\omega')\eta_{a'}
\cg^2_{a'}(\omega') \gamma_{a'}(\omega') \ q^A_{a'}.
\end{equation}
Unlike the Fermi liquid in the original works of Luttinger and Ward, 
the on-shell interaction vertex which represents the Fermi liquid is 
normalized by the appearance of intermediate spinon and holon states.

Incorporating the expression for the self energy derivatives within 
Eqs. (\ref{phasederivative}) and (\ref{susfromphase}), we can write 
the susceptibilities as 
\begin{equation}
\chi^A = \frac{{\rm Im}G_c}{\pi} 
\int \frac{d\omega}{2 \pi i} \sum_{\alpha \nu a}
q^A_{c\alpha \nu} \Gamma_{c \alpha \nu, a}(0,\omega) 
\cg^2_a(\omega) \eta_{a} q^A_a \gamma_a .
\label{chiA}
\end{equation}
Since we work in the wide band limit and $\gamma_c$ vanishes 
the summation over $\gamma_a \hat q_a$ gives non-vanishing 
contribution only for one particle species
\begin{eqnarray}
\chi^S &=&  \frac{{\rm Im}G_c}{\pi} K \sum_{\alpha \alpha'} 
{\tilde \alpha} {\tilde \alpha'}
 \int \frac{d\omega}{2 \pi i} 
 \Gamma_{c \alpha, b \alpha'}(0,\omega) 
\cg^2_b(\omega) \label{chiS}, \\
\chi^F &=&  \frac{{\rm Im}G_c}{\pi} N \sum_{\nu \nu'} 
{\tilde \nu} {\tilde \nu'}
\int \frac{d\omega}{2 \pi i} 
\Gamma_{c \nu, \chi \nu'}(0,\omega) 
\cg^2_{\chi}(\omega) ,\\
\chi^C &=& \frac{{\rm Im}G_c}{\pi} N \sum_{\nu \nu'}
\ \ \int \frac{d\omega}{2 \pi i} 
\Gamma_{c \nu, \chi \nu'}(0,\omega) 
\cg^2_{\chi}(\omega). 
\end{eqnarray}

%The Ward Identities.
%Luttinger Ward and Ward Identities in the presence of a spinon gap.
%Introducing the idea of the little $q^{a}$and big $Q^{a}$.
%
%Off-shell interactions. $\Lambda$ and $\Gamma$. 
%\underline{Stress the
%appearance of spinon and holon intermediate states.}
%Relationship between the susceptibilities and $\Gamma$. Interaction
%with external fields is carried mainly by spinons and holons.

\section{Low temperature Free energy}\label{}
The low temperature Free energy. Recapitulation of Luttinger Ward's
hard argument.  If we use the zero temperature Green's function
$\tilde{\cal
G}\equiv {\cal G}\vert_{T=0}$
, the leading change in the Free energy at finite temperatures
$\Delta F=F (T)-F (T=0)$
can be computed solely from the finite temperature
trace
of the Logarithm of the {\sl zero } temperature Green's function
\begin{eqnarray}\label{l}
\Delta F 
= \left[ 
T{\Str}\left\{\ln  \left(- \tilde{\cal G}^{-1} \right)
\right\} \right]^{T}_{T=0}.
\end{eqnarray}

This is a quite non-trivial result, and for clarity, here we repeat
the original Luttinger Ward derivation in the context of the current
work. 
At small finite temperature we have
two things to keep track of in the free energy, the discreteness of
the frequency summations, and the temperature dependence of the Green
functions. However, since the Free energy is stationary at low
temperatures with respect to small changes in ${\cal  G}$, to
calculate the low temperature corrections to the Free energy, we may
use the zero temperature Green function $\tilde{\cal
G}\equiv {\cal G}\vert_{T=0}$ in the Luttinger Ward expression
\begin{equation}\label{notsobigdeal}
F[T\sim 0] = T{\Str}\left[\ln  \left(- \tilde{\cal G}^{-1} \right)+
\tilde{\Sigma }
\tilde{\cal G} \right]+ Y[\tilde{\cal G},T],
\end{equation}
where $ ({\cal G}_{0}^{-1}- 
\tilde{\cal G}^{-1} 
)=\tilde{\Sigma }
$ is the zero temperature self-energy.

Small finite temperatures introduce  a discreteness into the frequency 
summations that produces corrections of order $T^{2}$ in the Free energy. 
To keep track of these changes we must write
\begin{equation*}
T\sum_{i \alpha_{n}}A[i\alpha_{n}] = \int_{-i\infty}^{i\infty} 
\frac{d\omega}{2 \pi i} 
[1 + \epsilon (\omega)] A(\omega) ,
\end{equation*}
where 
\begin{equation}\label{}
\epsilon (\omega) = (2 \pi i T \sum_{n}\delta (i\omega + \alpha_{n}) - 1)
\end{equation}
keeps track of the discreteness.  Luttinger and Ward made the
insightful observation that at low temperatures, 
one can regard $\epsilon(\omega)$ to be an infinitesimal, so that inside 
the Luttinger Ward functional, 
the leading order effect of finite temperature is taken into account
by making a variation $\tilde{\cal G}\rightarrow \tilde{\cal
G}+ \delta {\cal G}$ where 
\[
\delta {\cal G}= + \epsilon \tilde{\cal G},
\]
in other words, 
\begin{equation}\label{keyrel}
Y[\tilde{\cal G},T ] - \tilde{Y}[\tilde{\cg}]   = 
\int \frac{d\omega}{2\pi i} {\rm Tr}\left[ 
2\pi i \frac{\delta  \tilde{Y}}{\delta {\tilde{\cg}} (
\omega)}
\epsilon (\omega)\tilde{\cg} (\omega)\right],
\end{equation}
where $\tilde{Y}=Y[\tilde{\cg},0]$.
\begin{widetext}
It is instructive to briefly digress to 
see how this replacement works in the leading order
contribution to the Luttinger Ward functional, given by
\begin{equation}\label{}
Y_{1}[{\cal G},T]=  K\times N \times \left(\frac{V}{\sqrt{N}}
\right)^{2}T^{2}\sum_{\omega_{n},\nu_r}\cg_{c} (\omega_{n})\cg_{\chi} (i \nu_{r}-i\omega_{n}) \cg_{b} (i\nu_{r}).
\end{equation}
So we can write 
\begin{equation}\label{}
Y_{1}[{\tilde{G},T}]=  K V^{2} \int
\frac{d\omega_{1}d\omega_{2}d\omega_{3}}{(2\pi i)^{3}}
2 \pi i \delta (i\omega_{1}+i\omega_{2}-i\omega_{3})[1+\epsilon
(\omega_{1})]
[1+\epsilon (\omega_{2})][1+\epsilon (\omega_{3})]
\tilde{\cg}_{c} (\omega_{1})\tilde{\cg}_{\chi} ( \omega_{2})\tilde{\cg}_{b} (\omega_{3}).
\end{equation}
To leading order in $\epsilon$, we have
\begin{eqnarray}\label{l}
\delta Y_{1}&=&  K V^{2} \int
\frac{d\omega_{1}d\omega_{2}d\omega_{3}}{(2\pi i)^{3}}
2 \pi i \delta (i\omega_{1}+i\omega_{2}-i\omega_{3})
\bigl [\epsilon (\omega_{1})
+\epsilon (\omega_{2})+\epsilon (\omega_{3})\bigr ]
\tilde{\cg}_{c} (\omega_{1})\tilde{\cg}_{\chi} (\omega_{2})\tilde{\cg}_{b} 
(\omega_{3})\nonumber
\cr
&=&\int \frac{d\omega}{2\pi i} \left[ 
NK
 2\pi i \frac{\delta  \tilde{Y}_{1}}{\delta \tilde{\cg}_{c} (\omega)}
\epsilon (\omega)\tilde{\cg}_{c} (\omega)
+K 2 \pi i \frac{\delta  \tilde{Y}_{1}}{\delta \tilde{\cg}_{\chi} (\omega)}
\epsilon (\omega)\tilde{\cg}_{\chi } (\omega)
+N 2 \pi i \frac{\delta  \tilde{Y}_{1}}{\delta \tilde{\cg}_{b} (\omega)}
\epsilon (\omega)\tilde{\cg}_{b} (\omega)
\right]\nonumber
\cr
&=&\int \frac{d\omega}{2\pi i} {\rm Tr}\left[  
2 \pi i \frac{\delta  \tilde{Y}_{1}}{\delta {\tilde{\cg}} (\omega)}
\epsilon (\omega)\tilde{\cg} (\omega)\right],
\end{eqnarray}
where the factors of $N$, $NK$ and $K$ arise because the
functional derivative refers to the variation of $Y_{1}$ with respect
to a given spin/flavor leg of the propagator.
The same series of manipulations can be carried out on diagrams of
arbitrary order. 
\end{widetext}

Returning from the digression, let us now use the relation
\begin{equation}\label{}
\delta Y =  -\underline{Str}\left[\Sigma \delta {\cal G}
\right]\stackrel{T\rightarrow  0 }{= }-\int \frac{d\omega}{2 \pi i
}{\rm  Str} \left[\Sigma (\omega)\delta {\cal  G} (\omega)\right],
\end{equation}
to identify
\begin{equation}\label{}
2\pi i \frac{\delta  \tilde{Y}}{\delta \tilde{\cg} (\omega)} = -\eta 
\tilde{\Sigma}(\omega).
\end{equation}
With this identification, we can rewrite (\ref{keyrel}) as 
\begin{eqnarray}\label{l}
\Delta Y  &=& 
-\int \frac{d\omega}{2\pi i} {\rm Str}\left[ 
\tilde{\Sigma} (\omega)
\epsilon (\omega)\tilde{\cg} (\omega)\right]\cr
&=& 
-T\left[\underline{\rm Str}\left\{ 
\tilde{\Sigma }
\ \tilde{G} \right \}
\right]^{T}_{T=0}.
\end{eqnarray}
We see that the finite temperature 
change in the generating functional correction term exactly cancels
the change from the second,  self-energy
term in the Free energy functional, so that for small finite
temperatures
\begin{eqnarray}\label{l}
\Delta F &=& \left[ 
T{\Str}\left\{\ln  \left(- \tilde{\cal G}^{-1} \right)+
\tilde{\Sigma }
\tilde{\cal G} \right\} \right]^{T}_{T=0}+ \Delta Y\cr
&=& \left[ 
T{\Str}\left\{\ln  \left(- \tilde{\cal G}^{-1} \right)
\right\} \right]^{T}_{T=0},
\end{eqnarray}
as promised. 

Now in principle, this expression involves finite temperature
contributions from all fields, including the fractionalized
excitations. If we carry out the Matsubara summations, we see that
\begin{eqnarray}
\Delta F  = - \frac{T^2 \pi ^2}{3} \frac{1}{2\pi}
 \frac{d}{d \omega} \left[ {\rm Str} \ {\rm Im} \ \ln 
\left( - \tilde \cg^{-1}\right) \right]_{\omega=-i0^+}
\end{eqnarray}
and the density of states of each field  contributes to the specific
heat coefficient $\gamma = - d^2F/dT^2 $.  It is this 
observation that primarily motivates our gap hypothesis, for with 
a gap in the holon/spinon spectrum, the only remaining contribution 
to the linear specific heat derives from the conduction electrons
\begin{equation}
\gamma =  \frac{2 \pi^2}{3}
 \frac{d}{d \omega} \frac{1}{2\pi}
\left[ \sum_{\alpha \nu} \ {\rm Im} \ \ln 
\left( \tilde \cg_c^{-1} / \tilde (\cg_c^0)^{-1} \right) \right]
_{\omega=-i0^+},
\end{equation}
note that as was previously done for susceptibilities we subtract the 
effect of the empty conduction band to get the impurity contribution to 
the specific heat coefficient. We have removed the tilde sign $\tilde \cg
\rightarrow \cg$ as from now on we restrict our derivation to the Green's 
function at zero temperature. 

In the single impurity model, the low energy thermodynamics can then
be entirely expressed in terms of the conduction electron phase shift
\begin{equation}
\delta_c = {\rm Im} \ln \left[ 1 - \cg_c^0  \Sigma_c  \right],
\end{equation}
in terms of which
\begin{equation}
\gamma = \frac{ \pi^2}{3} \sum_{\alpha \nu} \ \frac{1}{\pi}
 \left. \frac{d \delta_c}{d \omega} 
\right|_{\omega = - i 0^+} .
\end{equation}
The linear coefficient of the specific heat is then
\begin{equation}
\gamma = \left. 
- \frac{ \pi^2}{3} \sum_{\alpha \nu} \frac{1}{\pi} {\rm Im} \cg_c 
\frac{d \Sigma_c}{d \omega} 
\right|_{-i0^+},
\label{gamma}
\end{equation}
and we see that the renormalization of the density of states is
entirely encoded in the frequency dependence of the conduction
electron self energy at the Fermi surface.

The derivative of the self energies with respect to the frequencies 
can be calculated in a similar way to the derivatives with respect 
to the fields $B^A$. The analogous expression is 
\begin{eqnarray}
\frac{d \Sigma_a(\omega)}{d \omega} = - \sum_{a'} 
\int \frac{d\omega '}{2 \pi i} 
\Gamma_{a a'}(\omega, \omega') \eta_{a'} \cg_{a'}^2(\omega')
\gamma_{a'}(\omega') \nonumber \\
+ \sum_{a'} \Gamma_{aa'}(\omega,0)  \eta_{a'}
\left. \frac{{\rm Im} \cg_{a'} }{\pi} \right|_{-i0^+}  .  
\label{dSdw}
\end{eqnarray}
The additional second term results from the discontinuous jump 
across the real axis: $\delta \cg = -2i{\rm Im} \cg(0-i\eta)$. In the first 
term the contribution from the local conduction electrons to the 
summation vanishes in the band limit. On the other hand in the second 
term only the conduction electrons contribute since the holon and boson 
are gapped and their spectral function vanishes 
at the Fermi level. Explicitly 
\begin{eqnarray}
\left. \frac{d \Sigma_c^{\alpha \nu}}{d\omega} \right| _{-i0^+} =
-\int \frac{d\omega }{2 \pi i} \left\{
\sum_{\alpha'} \Gamma_{c\alpha,b\alpha'}(0, \omega) \cg_{b}^2(\omega)
\ \ \ 
\right. \nonumber \\ \left.
-\sum_{ \nu'} \Gamma_{c\nu,\chi \nu'} (0,\omega) \cg_{\chi}^2(\omega) 
\right\} \nonumber \\
- \left. \frac{{\rm Im} \cg_{c} }{\pi}\right|_{-i0^+}
\sum_{\alpha' \nu'} \Gamma_{c\alpha\nu,c\alpha'\nu'}(0,0)  \ \ .
\label{dsdomega}
\end{eqnarray} 
\centerline {\includegraphics[width=3in]{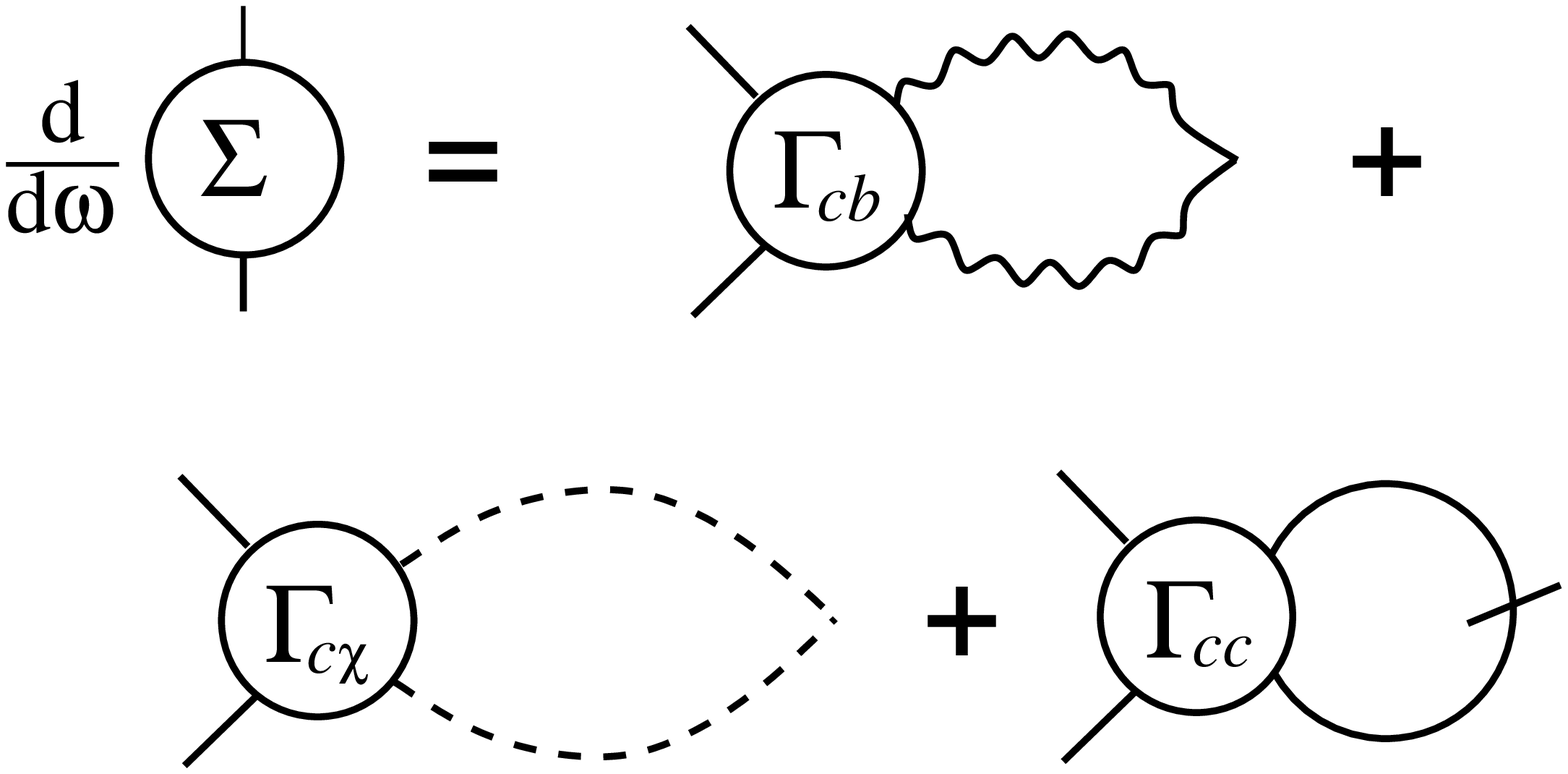}\label{dSdo}}

The next step in our derivation is to compute  the Ward identities 
that relate  the frequency and field dependences of
the electron self energies to the electron Landau parameters.

The derivation of the Friedel sum rules in Sec. 2  relied on the Ward 
identities of Eq.(\ref{Ward}) which were derived by shifting the 
frequencies in diagrams of the Luttinger Ward functional $Y[G]$ on 
by a shift $\delta \omega^A_a = \delta \omega q^A_a$ 
proportional to the conserved charge $Q^A$ carried by the corresponding 
particle: $G_c(\omega) \rightarrow G_c (\omega + \delta \omega^A_c)$.
In the Luttinger Ward diagrams such shifts correspond to shifts along 
along closed loops. Since the interaction vertex conserves the charges 
$Q^A$ it is also possible to think of similar shifts in the self energy 
diagrams\cite{yamada,yoshimori,mihalyzawadowski,yoshimorizawadowski}. 
The interesting case is when the external legs of the self 
energy carry the relevant charge $q^A_a \ne 0$, in this case in addition 
to the closed loops also the external frequency is shifted and 
the self energy change can be related to  
the frequency derivative of the self energy 
\begin{equation}
\frac{\delta \Sigma_a}{\delta \omega^A} = \frac{1}{q^A_a} 
\frac{d \Sigma_a}{d\omega}.
\end{equation}
These identities introduce in addition to Eq.(\ref{dSdw}) alternative 
equations for the frequency derivative of the self energies
\begin{eqnarray}
\frac{1}{q^A_a} \frac{d \Sigma_a(\omega)}{d \omega} = &-& \sum_{a'} 
\int \frac{d\omega '}{2 \pi i} 
\Gamma_{a a'}(\omega, \omega') \eta_{a'} \cg_{a'}^2(\omega')
\gamma_{a'} q^A_{a'} \nonumber \\
&+& \sum_{a'} \Gamma_{aa'}(\omega,0) q^A_{a'}  \eta_{a'} 
\left. \frac{{\rm Im} \cg_{a'} }{\pi} \right|_{-i0^+} . 
\label{deltasdeltaomega}
\end{eqnarray}
The conduction electrons carry the three charges $q^A_c \ne 0$, 
and it is possible to plug these derivatives into the expression 
for the specific heat coefficient of Eq.(\ref{gamma}), which gives
\begin{eqnarray}
\frac{\gamma}{\gamma_0} =  \frac{1}{NK} \left\{
\int \frac{d\omega}{2 \pi i} \sum_{\alpha \nu a}
q^A_{c\alpha \nu} \Gamma_{c \alpha \nu, a}(0,\omega) 
\cg^2_a(\omega) \eta_a \hat q^A_a  \gamma_a  \right. \nonumber \\
- \left.
\sum_{\alpha \nu a}
[\hat q^A_c]_{\alpha \nu} \Gamma_{c \alpha \nu,a}(0,0) q^A_a \eta_a 
 \frac{{\rm Im} \cg_{a} }{\pi} \right\}  ,\ \ \
\end{eqnarray}
with $\gamma_0 = NK \pi^2/3 \cdot {\rm Im} G_c/{\pi}$ the specific heat 
coefficient of the bare conduction band. Rewriting the latter 
equation we can relate the specific heat coefficient to the 
various susceptibilities
\begin{equation}
\frac{\gamma}{\gamma_0} = \frac{\chi^A}{\chi^A_0} -
\frac{1}{NK} \sum_{\alpha \nu a}
\hat q^A_{c\alpha \nu} \Gamma_{c \alpha \nu,a}(0,0) q^A_a \eta_a
 \frac{{\rm Im} \cg_{a} }{\pi}, 
\label{befw}
\end{equation}
where $\chi^A_0=NK {\rm Im} \cg_c/{\pi}$ are the bare Pauli 
susceptibilities. The latter expression can be simplified as a 
set of Ward identities relating the specific heat to the susceptibilities 
and the on-shell Fermi liquid interaction vertex
\begin{eqnarray}
\frac{\gamma}{\gamma_0} &=& 
\frac{\chi^S}{\chi^S_0} +
\frac{1}{NK} \frac{{\rm Im} \cg_c}{\pi} 
\sum_{\alpha \nu \alpha' \nu'} \tilde\alpha \tilde\alpha' \ 
\Gamma_{c \alpha \nu,c \alpha' \nu'}(0,0) \nonumber \\
&=& \frac{\chi^F}{\chi^F_0} +
\frac{1}{NK} \frac{{\rm Im} \cg_c}{\pi} 
\sum_{\alpha \alpha' \nu \nu'} \tilde\nu \tilde\nu'  \
\Gamma_{c \alpha \nu,c \alpha' \nu'}(0,0) \nonumber \\
&=& \frac{\chi^C}{\chi^C_0} +
\frac{1}{NK} \frac{{\rm Im} \cg_c}{\pi} 
\sum_{\alpha \alpha' \nu \nu'}  \ \ \ \ \
\Gamma_{c \alpha \nu,c \alpha' \nu'}(0,0). \nonumber \\
\label{Ward2}
\end{eqnarray}
In deriving these identities we assumed a gapped spinon and holon 
spectra which restrict the summation over $a$ in Eq.(\ref{befw}) 
to the conduction electrons. As discussed in the following section 
it is possible to directly relate the on-shell conduction electron 
interaction vertex $\Gamma_{c \alpha \nu, c \alpha' \nu'}$ to 
the Fermi liquid amplitudes $\phi_{\alpha \nu, \alpha' \nu'}$ by 
comparing the latter Ward identities to the Nozi\`eres Fermi liquid 
expressions. Note that it is possible to derive three 
additional spectral sum rules for the off-shell interaction vertices by 
equating Eq.(\ref{dsdomega}) with Eq.(\ref{deltasdeltaomega}).
 
\section{Yamada Yosida Relations}\label{}%  D. Yamada Yosida
As was mentioned in the introduction the phase shift in a Nozi\`eres 
type local Fermi liquid can be written as a function of the energy and 
the occupation 
\begin{equation}
\delta_{\alpha \nu}[ \omega, \{n_{k'\alpha',\nu'}\}]
= \delta_0 + \delta' \omega + \sum_{k' \alpha' \nu'} 
\phi_{\alpha\nu,\alpha'\nu'} \delta n_{k'\alpha'\nu'},
\end{equation}
where $\delta'$ is the energy derivative of phase shift and the impurity 
Landau parameter 
\begin{equation}
\phi_{\alpha \nu, \alpha' \nu'} = \langle \alpha \nu, \alpha' \nu'
| \hat \phi | \alpha \nu \alpha' \nu' \rangle
\end{equation} 
is the expectation value of the interaction between quasi particles.
The density of the quasiparticles is $\rho_0+\delta'/\pi$, where $\rho_0$ 
is the bare density of states, which implies that the impurity contribution 
to the specific heat coefficient is given by $\gamma = \gamma_0 
\delta'/\pi \rho_0$. 
For a local Fermi liquid the impurity Landau parameter can be parameterized 
in terms of two numbers $\phi_1$ and $\phi_2$ through
\begin{eqnarray}
\phi_{\alpha \nu,\alpha' \nu'} = \phi_1 \left( 1-\delta_{\alpha \alpha'}
\delta_{\nu\nu'} \right) + \phi_2 \left( \delta_{\nu\nu'}-
\delta_{\alpha\alpha'} \right).
\end{eqnarray} 
The response of the conserved charges $Q^A$ to their fields $B^A$ can be 
represented in terms of the phase shifts through
\begin{equation}
\chi^A = \frac{1}{\pi} \sum_{\alpha \nu} q^A_{c\alpha \nu} \frac{\Delta
 \delta_{\alpha \nu}}{ B^A} .
\end{equation}
The resulting susceptibilities read
\begin{eqnarray}
\frac{\chi^S}{\chi^S_0} - \frac{\gamma}{\gamma_0} &=& \sum_{\alpha \alpha'
\nu \nu'} \tilde \alpha \tilde \alpha'  \phi_{\alpha \nu,\alpha' \nu'}
= -\phi_1 - K \phi_2  \nonumber, \\ 
\frac{\chi^F}{\chi^F_0} - \frac{\gamma}{\gamma_0} &=& \sum_{\alpha \alpha'
\nu \nu'} \tilde \nu \tilde \nu' \phi_{\alpha \nu,\alpha' \nu'}
= -\phi_1 + N \phi_2  \nonumber ,\\
 \frac{\chi^C}{\chi^C_0} - \frac{\gamma}{\gamma_0} &=& \sum_{\alpha \alpha'
\nu \nu'} \phi_{\alpha \nu,\alpha' \nu'}
= (NK-1) \phi_1  + (N-K) \phi_2 \nonumber, \\ 
\label{FLsus}
\end{eqnarray}
where the Pauli susceptibilities are $\chi^A_0 = NK \rho_0/\pi$.
By comparing the latter equations to the set of equations (\ref{Ward2}), 
it is easy to read off the relations between the Landau parameters and 
the on-shell interaction vertex
\begin{equation}
\phi_{\alpha \nu,\alpha'\nu'} = -  \frac{{\rm Im} \cg_c}{\pi NK} \  
\Gamma_{c \alpha \nu,c \alpha' \nu'}(0,0).
\end{equation} 
From equations (\ref{FLsus}) it is possible to express the Fermi liquid 
parameters in terms of the reduced susceptibilities
\begin{eqnarray}
\phi_1 &=& \frac{\chi^C/\chi^C_0}{NK} -\frac{1}{N+K}
\left( \frac{\chi^F/\chi^F_0}{K} + \frac{\chi^S/\chi^S_0}{N} \right) 
\nonumber \\
\phi_2 &=& \frac{\chi^F/\chi^F_0 - \chi^S/\chi^S_0}{N+K},
\end{eqnarray}
which produce the following relation between the susceptibilities and specific 
heat coefficient
\begin{equation}
NK\frac{\gamma}{\gamma^0} = 
K\frac{N^2 -1}{N+K} \frac{\chi_s}{\chi_s^0} + 
N\frac{K^2-1}{K+N} \frac{\chi_f}{\chi_f^0} 
+\frac{\chi_c}{\chi_c^0},
\label{YY}
\end{equation}
generalizing the Yamada Yosida relation\cite{yamada,yoshimori,JAZ98} to 
our infinite-U model with $K$ channels.  
Note that also the on-shell interaction vertex can be parameterized in 
a similar way
\begin{equation}
\Gamma_{c \alpha \nu,\alpha'\nu'}(0,0) = \Gamma_1 (1 -\delta_{\alpha \alpha'}
\delta_{\nu\nu'})+\Gamma_2 (\delta_{\nu\nu'}-\delta_{\alpha\alpha'}),
\end{equation}
with $\Gamma_{1,2} = -\pi NK  \phi_{1,2} /{\rm Im} \cg_c $. These two Fermi 
liquid parameters control the low temperature thermodynamics, and are 
renormalized due to intermediate spinon and holon states. 

An interesting observation is that the prefactors of the susceptibility 
terms in Eq.(\ref{YY}) correspond to the three central charges of the 
$\widehat{SU(N)_K}$, $\widehat{SU(K)_N}$ and $\widehat{U(1)}$ Kac-Moody 
algebras respectively\cite{TsvelikBook}.  
In terms of the relevant Kac-Moody currents the free fermion problem 
decomposes to the free current Hamiltonians ${\cal H}^0_S$, ${\cal H}^0_F$ 
and ${\cal H}^0_C$. The fraction of the degrees freedom in section $A$, 
$\gamma_A^0/\gamma^0$, is given by the ratio of the central charge 
of section $A$ to the total central charge $NK$\cite{cardy}. 
Equation (\ref{YY}) emerges 
under the assumption that for the fully screened impurity problem the fixed 
point Hamiltonians have a similar form to the bare Hamiltonians with 
renormalized Fermi velocities and no residual degrees of freedom\cite{Affleck95}
\begin{equation}
{\cal H}^{FP}_A = \left( 1+ \frac{3\pi \lambda_A}{\ell} \right) {\cal H}_A^0, 
\end{equation}  
where $\lambda_A$ represents the impurity effective couplings and $\ell$ is 
the size of the system. Under this assumption $\chi_A/\chi_A^0 = \gamma_A/ 
\gamma_A^0 = 3\pi \lambda_A/\ell$, where $\gamma_A$ is the impurity contribution 
to the specific heat coefficient in section $A$. Summing up the impurity 
contribution to the specific heat coefficient $\gamma=\sum_A \gamma_A$ 
we recover Eq.(\ref{YY})
\begin{eqnarray}
\frac{\gamma}{\gamma^0} = \sum_A \frac{\gamma_A^0}{\gamma^0} 
\frac{\chi_A}{\chi_A^0} = \ \ \ \ \ \ \ \ \ \ \ \ \ \ \ \ \ \ \ \ \ \ 
\ \ \ \ \ \ \ \ \ \ \ \ \ \ \ \ \ \ \ \ \  \nonumber \\ 
=\frac{1}{NK} \left[ K\frac{N^2-1}{N+K} \frac{\chi_S}{\chi_S^0}
+ N\frac{K^2-1}{N+K} \frac{\chi_F}{\chi_F^0} + \frac{\chi_C}{\chi_C^0} 
\right] . 
\nonumber 
\end{eqnarray} 

%  
%Relationship between Ward Identities and the Nozi\`eres
% picture.
%\underline{Stress the
%appearance of spinon and holon intermediate states inside the Fermi
%liquid interaction.}
%
%Cancellation identity for the on-site interactions, and resulting identity.
%
%Interpretation in terms of conformal charges, $\gamma_{a}= C_{a}\chi_{a}$.

\section{Shiba -Korringa relations}\label{Shiba}
% Large bandwidth limit - spin vertex is entirely the local vertex.

The Korringa-Shiba relations relate the impurity's nuclear-spin -- lattice 
relaxation time $T_1$ with the Knight shift $K$ at small magnetic fields. 
Alternatively it relates also between the impurity static susceptibility 
and the low frequency impurity spin power spectrum. In the context of our 
paper this relation is important as a manifestation of the soft modes 
formation in the susceptibility power spectrum even though the spinon 
spectrum is gapped.  
This relation was first deduced from experimental data by Korringa, and 
only later it was proved\cite{shiba} by Shiba for the finite-U Anderson 
model summing up the perturbative expansion in U to arbitrary order. Below 
we show how Shiba's arguments can be extended to the infinite-U Anderson 
model and to the Kondo model by summing up the perturbative expansion in 
$1/N$. 

The expression deduced by Korringa reads
\begin{equation}
K^2 T_1 T = C,\label{Korringa}
\end{equation}
where $C$ is a constant. The Knight shift $K$ describes the mean field 
response observed by the nuclear moment, which for low Larmor frequency is
proportional to the static impurity susceptibility $\chi_S$. On the other 
hand the relaxation rate $T_1^{-1}$ is proportional to the low frequency 
power spectrum of the impurity susceptibility. Explicitly in terms of 
impurity susceptibilities $K$ and $T_1^{-1}$ are given by
\begin{eqnarray}
K &=& A_{\rm HF} \  {\chi_S^{zz}}'(\Omega) \nonumber, \\
T_1^{-1} &=& - T (g_N \mu_N)^2 A_{\rm HF}^2 \  
\frac{ {\chi^{+-}_S}''(\Omega-i0^+)}{\Omega} 
, \nonumber 
\end{eqnarray}
where $A_{\rm HF}$ is the hyperfine coupling and $g_N$ and $\mu_N$ are 
the nuclear spin g-factor and Bohr magneton respectively, and $\Omega$ is 
the Larmor frequency.
The latter expression with Korringa's empirical rule give immediately 
the Shiba relation between the static susceptibility and the spin power 
spectrum at low frequencies where the susceptibilities become isotropic
\begin{equation}
\left. \frac{\chi_S''(\Omega-i0^+)}{\Omega} \right|_{\Omega 
\rightarrow 0} = - \frac{\chi_S^2}{C (g_N \mu_N)^2}.
\label{shibar1}
\end{equation} 

Extending Shiba's arguments to the infinite-U Anderson model we 
see that by taking the derivative of the dynamic susceptibility 
$d\chi'' /d\Omega$ and substituting $\Omega \rightarrow 0$, the only 
contribution to the low frequency power spectrum comes from the 
following diagram

\begin{equation}
\centerline {\includegraphics[width=2.5in]{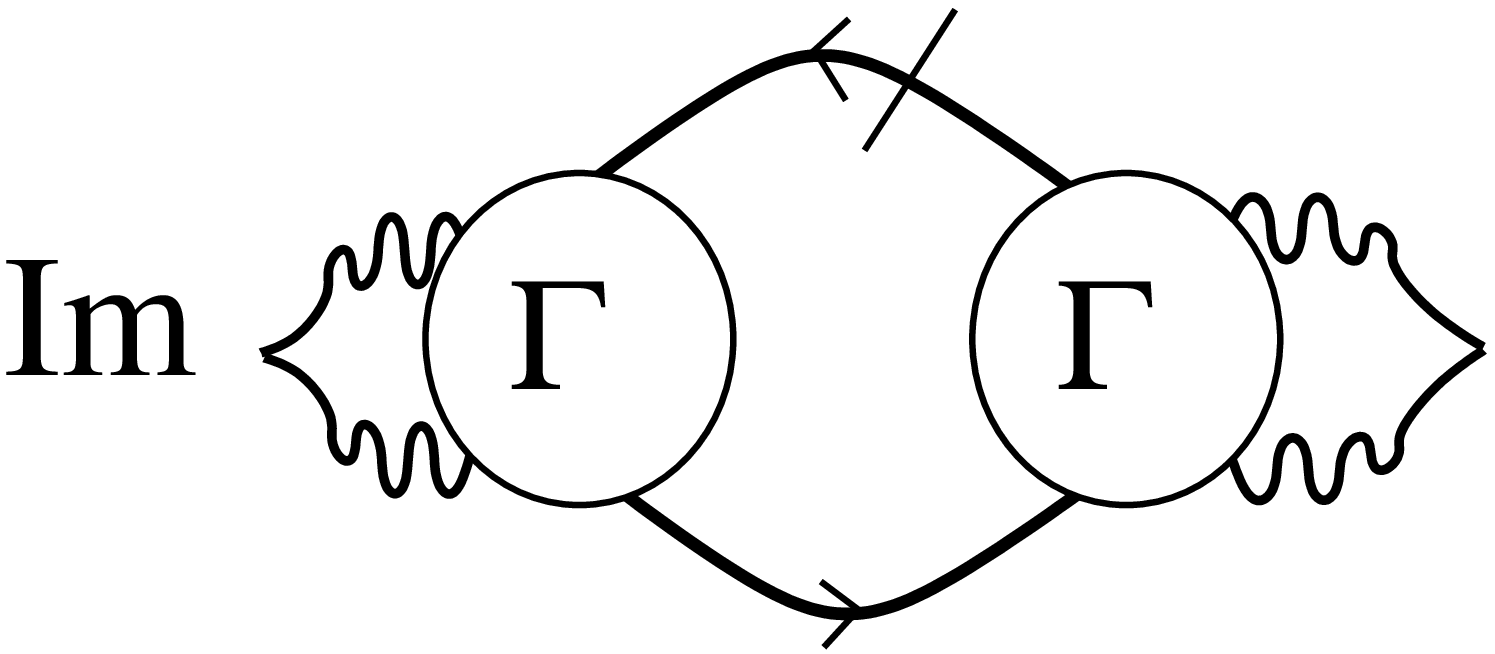}}.
\label{shibad}\nonumber
\end{equation}

\noindent The line that 
intersects the upper conduction electron line denotes the 
non-analytic part of the derivative of the green function which arises 
from the jump across the real axis $ d \cg_c/d\omega |_{\rm na} = -2 i  
\cg_c''(-i0^+) \delta(i\omega)$. Since the boson and holons are gapped 
there are no similar contributions from derivatives of the boson and holon green 
functions. The analytic part of the green function derivatives 
produce no imaginary contribution and should not be considered here. 
The other diagrams arise from derivatives of bare interaction vertex 
$\Lambda$. These diagrams have more than one electron-hole pair hence 
their contribution to the imaginary part is super linear in 
$\Omega$ and the derivatives vanish as $\Omega \rightarrow 0$ \cite{shiba}. 

A straightforward calculation of the diagram (\ref{shibad}) gives
\begin{eqnarray}
\left. \frac{\chi_S'' (\Omega-i0^+)}{\Omega} 
\right|_{0} 
= \frac{K (\cg_c''(-i0^+))^2}{\pi}  
\sum_{\alpha \alpha' \alpha''} \tilde \alpha \tilde \alpha' \times \nonumber \\
\int_{-i\infty}^{i\infty} \frac{d \omega d\omega'}{(2\pi i)^2} 
\cg_b^2(\omega) \Gamma_{b\alpha c \alpha''} (\omega,0)  
\Gamma_{c \alpha'' b\alpha'}(0,\omega') \cg_b^2(\omega') \nonumber .\\
\label{power1}
\end{eqnarray}

In order to obtain the relation between the power spectrum of Eq.(\ref{power1}) 
and the static susceptibility of Eq.(\ref{chiS}) we need to resort to 
the symmetry properties of the vertex function $\Gamma$. In addition to 
the symmetry to the exchange of variables which is inherited from the 
bare interaction vertex $\Lambda$, there is an additional symmetry of the 
vertex function as a two particle vertex. In general a two particle vertex 
$\Gamma_{\alpha \bar \alpha, \alpha' \bar \alpha'}$ is a function of 
the incoming spins $\alpha$ and $\alpha'$ and of the outgoing spins
$\bar \alpha$ and $\bar \alpha'$. In our model there is no annihilation 
or creation of spin therefore each incoming spin must identify with an 
outgoing spin
\begin{equation}
\Gamma_{\alpha \bar \alpha, \alpha' \bar \alpha'}= \Gamma^{(1)} 
\delta_{\alpha \bar \alpha} \delta_{\alpha' \bar \alpha'}
+ \Gamma^{(2)} 
\delta_{\alpha \bar \alpha'} \delta_{\alpha' \bar \alpha}.
\end{equation} 
In the full interaction vertex function the incoming and outgoing 
spins of each particle are already identified $\alpha = \bar \alpha$ 
and $\alpha' = \bar \alpha'$, hence we can write
\begin{eqnarray}\label{l}
\Gamma_{c\alpha b\alpha'} (\omega,\omega') =
\Gamma^{(1)}_{cb}(\omega,\omega') + \Gamma^{(2)}_{cb}(\omega,\omega')
\delta_{\alpha \alpha'}. 
\end{eqnarray}
The static susceptibility of Eq.(\ref{chiS}) gets a contribution only 
from the exchange part of the vertex $\Gamma^{(2)}$ as the direct part 
$\Gamma^{(1)}$ drops out of the summation $\sum \tilde \alpha \tilde 
\alpha' \Gamma$. The resulting simplified expression is 
\begin{equation}
\chi_S = NK \frac{\cg_c''(-i0^+)}{\pi} \int_{-i\infty}^{i\infty}
\frac{d\omega}{2 \pi i}
\cg_b^2(\omega) \Gamma^{(2)}_{cb}(0,\omega).
\label{chiSs}
\end{equation}

A similar simplification can be presented for the power spectrum of 
Eq.(\ref{power1}) by noticing that 
\begin{eqnarray}
\sum_{\alpha} \tilde \alpha \Gamma_{b\alpha c\alpha''} &=& \tilde \alpha'' 
\Gamma^{(2)}_{bc} , \\
\sum_{\alpha'} \tilde \alpha' \Gamma_{c\alpha'' b\alpha'} &=& \tilde \alpha'' 
\Gamma^{(2)}_{cb}.
\end{eqnarray} 
Taking further advantage of the symmetry to variables exchange we can 
rewrite the power spectrum as 
\begin{eqnarray}
\left. \frac{\chi_S''(\Omega-i0^+)}{\Omega} \right|_{0}
= \frac{NK(\cg_c'')^2}{\pi}
\left( \int \frac{d\omega}{2\pi i} \cg_b^2(\omega) \Gamma^{(2)}_{cb}(0,\omega)
\right)^2.
\nonumber
\end{eqnarray}
The right-hand side of the latter equation can be expressed in term of the 
static susceptibility of Eq.(\ref{chiSs}) which gives the Shiba relation
\begin{equation}
\left. \frac{\chi_S''(\Omega-i0^+)}{\Omega} \right|_{0} = \frac{\pi}{NK} 
\chi_S^2.
\end{equation}
According to this derivation the Korringa constant $C$ of Eq.(\ref{Korringa}) 
is $NK / [\pi(g_N \mu_N)^2]$.

This result implies that the 
dynamical susceptibility is dominated at low frequencies by intermediate
electron-hole pair states. These intermediate states  of the Fermi liquid
dress the dynamical susceptibility 
to produce a  non-vanishing power spectrum at low frequencies.
While the spin-charge decoupled excitations remain gapped, 
the are soft magnetic modes that are carried by intermediate quasiparticles.

%\section{Model calculation of the dynamical susceptibility of the
%Kondo model}\label{}
%
%Emphasizing the last point.

%The special case of the leading approximation. 
%  Our  calculation of Chi" /omega
%  
\section{Discussion}

This paper was motivated by an interest in the
way a Landau Fermi liquid can emerge within the  constrained
Hilbert space of highly correlated electron systems. 
In such highly constrained systems, we can no longer appeal to adiabaticity
or the use of infinite order expansions in the strength of the interaction.
By combining the Luttinger Ward approach with a new class of large $N$
expansion for the infinite $U$ impurity Anderson model, we have obtained
a novel  perspective on the Fermi liquid
that may have implications for our understanding of the
corresponding Anderson and Kondo lattice models. 

In the single impurity model, we find
that the Fermi liquid  contains 
low lying electronic quasi particles that break-up
into fractionalized excitations above a certain gap
energy $\Delta_{g}$. 
The assumption
that this gap holds to all orders in our analysis leads to 
a full set of conserving relationships between dynamic and
thermodynamic variables, but it also has qualitative implications for
our understanding of the Landau Fermi liquid.

Indeed, our results indicate a rather intimate relation 
between the opening of a gap in the spinon and holon spectra,
and the formation of a heavy Fermi liquid.  
In particular, the Landau scattering
amplitudes that appear in our treatment contain, as intermediate states,
the virtual excitation of spinon or holon pairs, for example, 
\centerline {\includegraphics[width=3.5in]{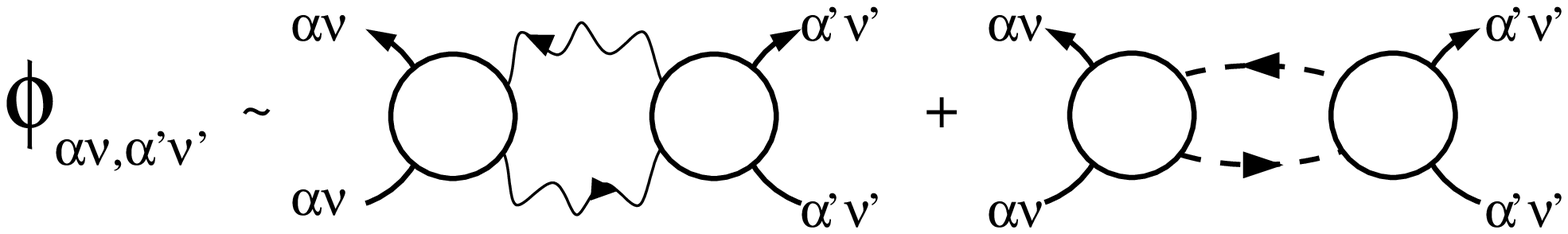}\label{Landau}}. 
These amplitudes are proportional to the inverse of the gap, 
and are finite so long as the gap is present. 
Through this description we showed how these virtual excitations 
give rise to the renormalization of the electron-electron interaction, 
and a whole host of conserving relationships known to be valid for the
Fermi liquid. 

But is this really a Fermi liquid? Landau's notion of a Fermi
liquid certainly has no place for fractional quasiparticle
excitations, even if those
particles are gapped. On the other hand, 
the fractional particles that appear in this theory involve fields
with matrix elements that link states of
{\sl different} $Q$: states that corresponding to 
models with different impurity spin $S=Q/2$. The excitation ``gap''
in the spinon spectrum involves
excitations from the ground-state of the fully screened 
model with spin $S_{0}=K/2$, to excited states of the under-screened
model, with $S=S_{0}+\frac{1}{2}$, 
\begin{eqnarray}\label{l}
G_{b} (t)&\sim& \sum_{\alpha }\langle S_{0}+\frac{1}{2},\alpha \vert b_{\alpha }\dg
\vert  S_{0}\rangle e^{- i\bigl (E_{\alpha }[S_{0}+\frac{1}{2}]
-E_{g}[S_{0}]\bigr )t}\cr
&\sim& e^{- i\bigl(E_{g} [S_{0}+\frac{1}{2}]-E_{g}[S_{0}]\bigr)t }.
\end{eqnarray}
where  $E_{g}[S_{0}]$ and $E_{g}[S_{0}+\frac{1}{2}]$ are the ground-state energies of the
fully screened and under-screened models respectively.  Thus the
appearance of the gap in the spinon/holon spectrum is purely a
consequence of the stability of the fully screened ground-state.

Physical quantities, such as the spin or charge correlation
functions, or the f-spectral function 
involve combinations of spinons or holons which do not
change $Q$ and do not display this gap. 
For example, 
the spin raising operator creates a spinon/antispinon pair.  
Such excitations are ``gauge neutral'' and are 
contained in the physical Hilbert space of definite $Q$. These pair excitations
couple to the conduction sea, converting the sharp gap of the infinite
$N$ limit into a pseudogap at finite $N$. 
The associated time-scale in this gap
$t_{c}\sim \hbar /\Delta_{g}$ gap can be interpreted as a 
confinement time-scale, for after this period the spinon-antispinon pair
combines with holons from the sea
to produce an electron-hole pair
\begin{equation}\label{}
s_{\alpha }+ \bar s_{\beta
}\stackrel{t\gtappr\frac{\hbar}{\Delta_{g}} }{\longrightarrow }
e_{\alpha \mu }+ \bar {e}_{\beta \mu}.
\end{equation}
Diagrammatically, 
\begin{equation}\label{props}
\centerline {\includegraphics[width=3.in]{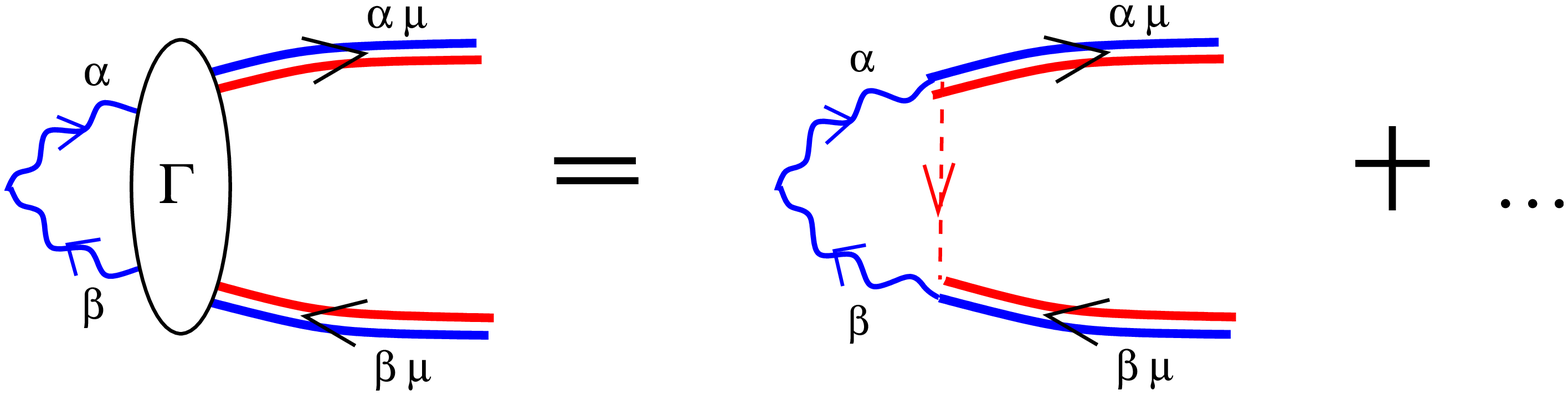}}
\end{equation}
(This process can be loosely compared to 
``quark jets'' in a collider, where quark-antiquark pairs
rapidly combine with virtual quarks from the vacuum, to form 
mesons or baryons.)

The processes that confine the
spinons are precisely those that gave rise to the gapless spin
spectral function.
From our derivation of the Shiba relationship,  we know that the decay of  spinon-pairs
into the electron-hole continuum is constrained to give rise to a  spectral function
with a pseudo-gap whose zero-frequency intercept is determined 
by the relation $\chi '' (\omega)/\omega\vert_{\omega=0}= \chi^{2}
\pi/ (NK)$: 
  \centerline {
\includegraphics[width=3.1in]
{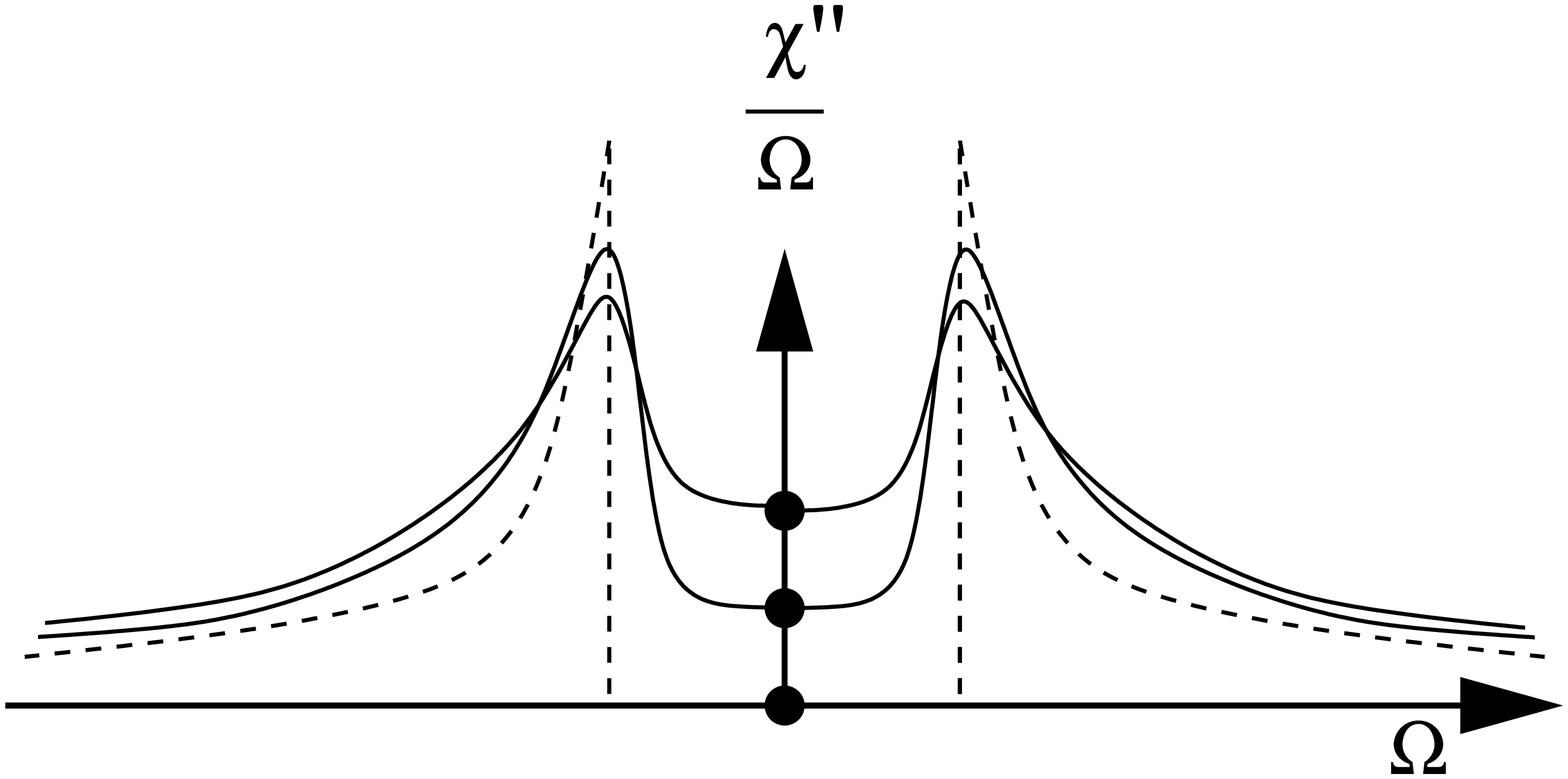}\label{powerspectrum}
}
where the dots will represent the DC limit $\chi ''
(\omega)/\omega\rightarrow \chi^{2} \pi/ (NK)
$
predicted by the Shiba
and the dashed line represents  the gapped, $N 
\rightarrow \infty$ limit.
A detailed calculation of this conserving spectrum within the leading
Kadanoff Baym approximation\cite{lebanon}  would be an interesting exercise for
future work. 
The Fourier transform of this power 
spectrum leads to the correct $1/t^{2}$ decay of the spin correlation 
function $\langle S (t)S(0)\rangle $. 
Thus the confinement of the
spinons through their coupling to the electron continuum is an
integral part of the emergence of Fermi liquid behavior in the single
impurity model. The pseudo-gap in this function is quite unremarkable
from the Fermi liquid perspective, where the 
the  offset peaks in the dynamical susceptibility are a well-known signature of
the asymmetric Kondo resonance
. 

In this way, the confinement of spinons and holons to form a Landau
Fermi liquid, appears very naturally in the Schwinger boson approach
to the single impurity, and it can be seen as a direct consequence of the way the gapped spin
fluid couples to the conduction fluid.  
This is a very different kind of confinement to that considered in the
context of insulating spin liquids, where there is a broken gauge
symmetry and topological tunneling
effects of the gauge field are required to understand the confinement
mechanism\cite{deconfinedcrita,senthil,hermele}.

\subsection{Lattice: heavy fermions}

There are some interesting implications of this kind of picture
for the more general lattice models. Consider for example, the (K-channel)
infinite U Anderson-Heisenberg lattice model 
\[
{\cal H}_{Lattice} = \sum_{{\vec k}\nu \alpha} \epsilon_{\vec k}
c^{\dagger}_{{\vec k}\nu \alpha } c_{{\vec k}\nu \alpha}+\sum H_{I} (j) 
+ \sum_{<i,j>} {\cal H}_{m}(i,j).
\]
Here 
\[
H_{I} (j)  =
\frac{{V}}{\sqrt{N}}\sum_{\nu \alpha}
\left[ \psi \dg_{j\nu\alpha }
\chi^{\dagger}_{j\nu} 
b_{j\alpha}
+{\rm H.c} \right] \nonumber 
+ \epsilon_0 \sum_{\alpha} b^{\dagger}_{j\alpha} b_{j\alpha} ,
\]
describes 
the atomic orbitals at the different sites and their 
hybridization with the conduction band,
where $\psi \dg_{j\nu\alpha }=\sum_{\vec{k}}c^{\dagger}_{{\vec k}\nu \alpha} e^{-i \vec{ k}\cdot
\vec{R}_{j}}$ creates a conduction electron at site $j$. The final 
terms  ${\cal H}_{m} (i,j)$ describes the antiferromagnetic 
spin interaction between nearest neighbors, where
\[
{\cal H}_{m}(i,j) 
= -\frac{J_{H}}{N}B\dg _{ij}B_{ij}.
\]
and $B\dg _{ij}= \sum_{\alpha }{\rm  sgn} (\alpha)
b\dg _{i\alpha }b\dg _{j-\alpha }
$ creates a singlet pair of bosons along the bond $(i,j)$.
In the lattice model, the conserved charges $Q_{j}= n_{b} (j)+ n_{\chi } (j)$
replace the no-double occupancy constraint at each site, and as
before, the fully screened case corresponds to $Q_{j}=K$

%\cite{noteNeel}. 
The manipulations that we have carried out in this paper can
be extended to the model, but certain caveats apply. 
One of key differences in the lattice model, is that in the large $N$
limit the
antiferromagnetic interactions induce {\sl pair condensation}
($- (J_{H}/N)\langle B_{ij}\rangle = \Delta_{ij}$) of the
Schwinger bosons, which mark the onset of short-range magnetic
correlations\cite{AA88}.  
The important  effect of the 
antiferromagnetism at a mean-field level is described by 
\begin{equation}
{\cal H}_{m}(i,j) 
\rightarrow 
\left\{ \Delta_{ij} B\dg _{ij}
+\bar \Delta_{ij} B_{ij}
+
N\frac{\bar \Delta_{ij} \Delta_{ij}}{J} \right\}
\label{latticeMF}
\end{equation} 
The appearance of a spinon-pair condensate is restricted to regions of
large $J_{H}\gtappr T_{K}$.  
Once pair condensation occurs, the 
local gauge symmetry associated with conservation of the $Q_i = K$ is
broken, replaced by a global 
$U(1)$ symmetry (where the difference of the $Q_{i}$ on even and odd
sublattices is conserved). Now, individual spinon and holon excitations 
start to propagate between sites. The phase of the bond-variables
becomes elevated into a compact $U (1)$ gauge theory\cite{hermele}.

We now discuss the qualitative phase diagram of this model, using some
of the insights gained from our impurity treatment.
A rigorous application of our methods to the
lattice requires that we include the $U (1)$ gauge fields into the Luttinger Ward
functional, a task that is beyond the scope of the current discussion.
However, we expect that many qualitative aspects of the current
analysis will survive.

There are two variables that tune quantum fluctuations in the lattice
model,  (Fig. \ref{lattice_pd} )
\begin{itemize}
\item ``Spin fluctuation parameter'' $1/{\tilde{S}} =\frac{N}{2S}$. 

\item ``Doniach parameter'' $t=T_{K}/J_{H}$ where $T_{K}$ is the impurity Kondo
temperature. 

\end{itemize}
The first parameter tunes the strength of the quantum spin
fluctuations and is loosely equivalent to the effect of lattice frustration.
By tuning the value of $1/{\tilde{S}}$, we can tune from classical 
antiferromagnetism, at small $1/\tilde{S}$ where the Schwinger boson individually condenses to form
an antiferromagnet, to a quantum antiferromagnetism at
large $1/\tilde{S}$, where a spin-liquid with a gap develops.  

\vspace{0.2in}
\fg{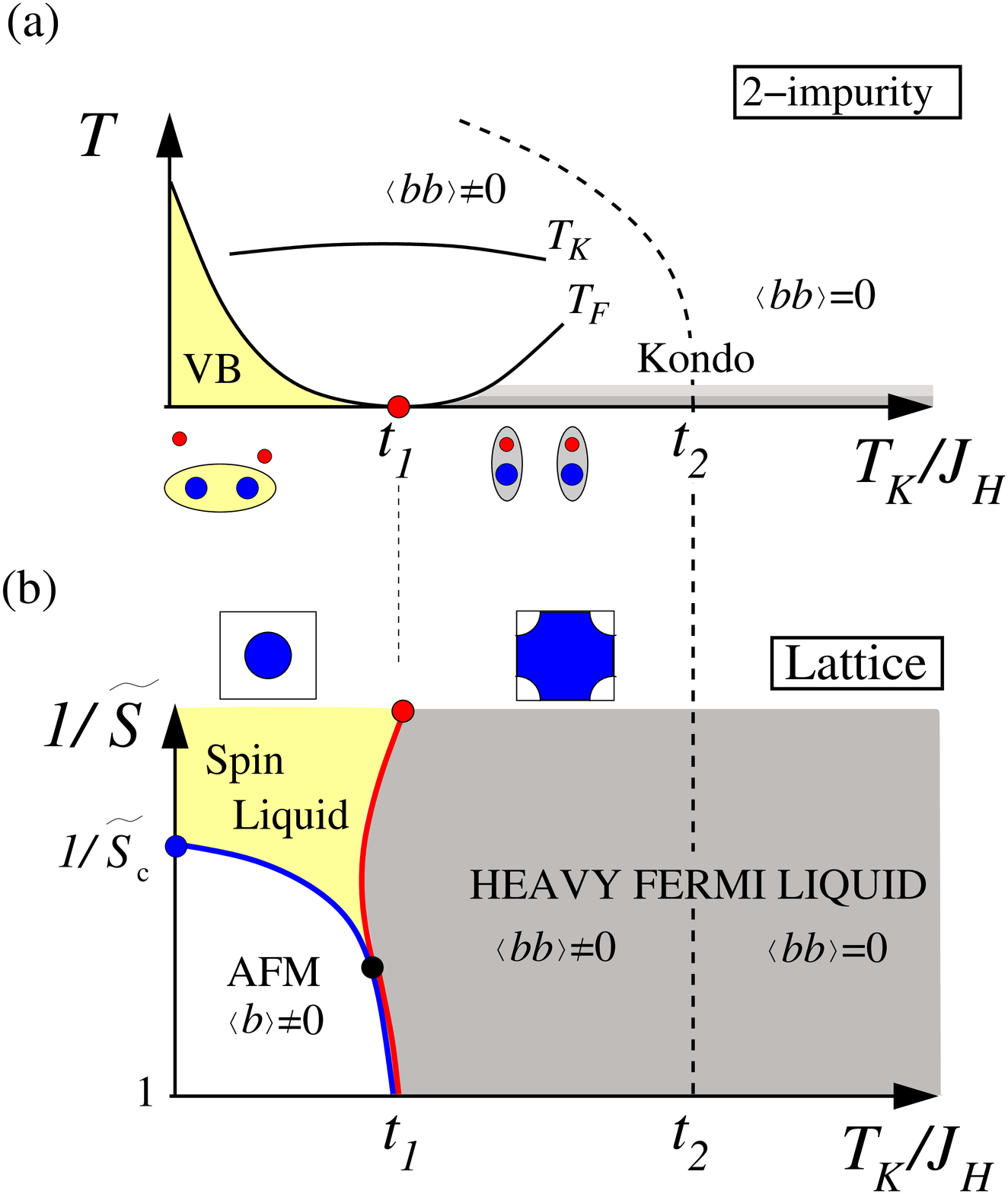}{lattice_pd}
{Showing the large $N$ phase diagram of two impurity and lattice
models as a function of the ratio $t=T_{K}/J_{H}$.
Boson pair condensation $\langle bb \rangle \ne 0$ 
develops for $t<t_{2}$ inducing  short-range magnetic
correlations that compete with the ``Kondo'' screening. 
(a) Computed large $N$ phase diagram of the two impurity Kondo problem
\cite{rech} displaying the relative temperature scales. 
$T=T_{K}$ and $T=T_{F}$ denotes the lower and upper scales associated
with the two stage screening process\cite{rech}. The Fermi temperature
and spinon gap close at the Varma Jones quantum critical point at 
$t=t_{1}$, which 
separates the valence bond state from the two impurity Kondo state. 
(b) Speculative sketch of the zero temperature phase diagram for the
Kondo lattice, based on the discussion in the text, 
showing dependence on the spin fluctuation parameter $1/\tilde{S}= N/ (2S)$. 
At large $1/\tilde{S}$ and small $t$, a spin liquid (SL) forms. 
At small $1/\tilde{S}$ and small $t=T_{K}/J_{H}$, the Schwinger boson
pair condenses to produce an antiferromagnet.  The
spin-liquid/AFM
 and spin-liquid/HF\cite{vojta,senthil} phase transition are expected
 to merge into a single quantum critical
point that separates the heavy electron state from the
antiferromagnet. The spinon and holon gaps close at each of these
quantum phase transitions.}

The second parameter tunes the strength of the Kondo effect. When $t$
is small, boson pair condensation takes place and the spin physics is that of an insulating
antiferromagnet or spin liquid with a decoupled background of conduction
electrons. At large $t>t_2 \sim 1$, no pair condensation occurs preserving the locality 
of the spinon and holon propagators. The direct extension of our methods to this 
case predicts a ``local'' heavy electron state which develops a large Fermi 
surface volume and enhanced momentum independent scattering amplitudes.
In the single impurity model, the charge of the localized state is given
by the difference of the electron and the holon phase shifts
Eq.(\ref{friedelsr})
\[
K-n_{\chi }=NK\frac{\delta_{c }}{\pi} 
- K\frac{\delta_{\chi }}{\pi}.
\]
The corresponding Luttinger theorem for the 
lattice\cite{indranil} tells us that the charge density 
is given by the difference of the conduction electron and holon Fermi
surface volumes
\[
n_{e}= NK \frac{{\rm v}_{FS}}{(2\pi)^{D}} - K\frac{\rm v_{\chi }}{(2 \pi)^{D}}.
\]
This expression holds some interest for us, because in the spin liquid phases
of the model, where the Kondo effect is inactive, we expect the Fermi
surface volume of the holons to be zero, giving a small Fermi surface
for the electrons. By contrast, at large $t=T_{K}/J_{H}$, where the
Kondo effect occurs, the Fermi surface volume of the gapped holons is
now at a maximum $v_{\chi }/ (2\pi)^{D}=1$, 
causing the electron Fermi surface to expand by an amount $\Delta
v_{FS}= 1/N$, forming a large Fermi surface. So long as
the holons preserve their gap, the volume of these Fermi surfaces will remain
fixed.

The effect of tuning the ratio $t= T_{K}/J_{H} $ has already been
considered in the context of a two impurity Kondo model\cite{rech}
(Fig. 3.(a)).  Fig. 3 (b) shows the schematic behavior expected in the
large $N$ lattice model.  Here, depending on the size of
$1/\tilde{S}$, either a spin liquid (large $1/\tilde{S}$) or an
antiferromagnet (small $1/\tilde{S}$) develops at
small Kondo coupling $t=T_{K}/J_{H}$.  At large t however, the heavy
fermi liquid, with a gap to spinons and holons will remain stable,
with Fermi liquid scattering amplitudes that are determined by the
exchange of virtual spinons and holons.
Provided the phase transition between these two limits is second
order, then the spinon-holon gap must close at the transition line,
with the concomitant divergence of the Landau scattering amplitudes. 

The closing  of the spinon gap may have important implications for the
evolution of the Fermi liquid parameters near the antiferromagnetic
instability. 
In the Schwinger boson scheme, 
at the antiferromagnetic instability, the spinons  condense at one
half the magnetic wave-vector, becoming gapless at two points in
momentum space $\vec{q}= \pm \vec{Q_{0}}/2$.  This will give rise to
strong electron -electron scattering at 
\begin{equation}\label{}
\vec{ q}= \vec{Q}_{0}/2\pm
\vec{Q}_{0}/2=\left\{
\begin{array}{c}
0,\cr
\vec{Q}_{0},
\end{array} \right.
\end{equation}
as illustrated in 

\centerline {
\includegraphics[width=1in]
{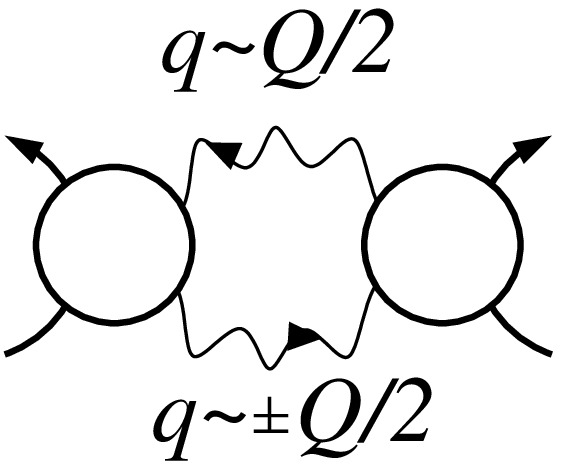}\label{bscorr}.
}
The appearance of strong scattering at a staggered wave-vector is
well-known in a spin-density wave scenario. However, the additional appearance of
strong scattering at $\vec{q}=0$, is a new feature.  In other words,
a subleading effect of the antiferromagnetic instability driven by
spinon condensation, is the development of strong forward scattering
reminiscent of an almost {\sl ferromagnetic} metal.  Features of this
sort have been observed near the field-induced quantum critical point
in $YbRh_{2}Si_{2}$\cite{gegenfm}.

These qualitative features of the large $N$ limit of our model
are a topic of current active examination, and we hope to report on
them in greater detail in forthcoming work. 

This work was supported by DOE grant number DE-FE02-00ER45790.
We are particularly indebted to Jerome Rech, for discussions related
to this work and for sharing with us his 
calculations of the Yoshida-Yoshimori identity in the Kondo limit. 
We should like to thank M.Fisher, C. Pepin, T. Senthil, Q. Si  and A. Vishwanath
for interesting discussions related to this work. 
P. C would like to thank the Aspen Center for Physics and
the Lorentz Center, Leiden, where part of this work was carried out. 
%  
% 
%-------------------

\end{document}